\documentclass[journal,10pt]{IEEEtran}

\pdfoutput=1

\makeatletter
\def\ps@headings{%
\def\@oddhead{\mbox{}\scriptsize\rightmark \hfil \thepage}%
\def\@evenhead{\scriptsize\thepage \hfil \leftmark\mbox{}}%
\def\@oddfoot{}%
\def\@evenfoot{}}
\makeatother
\usepackage{booktabs}

\usepackage[tight,footnotesize]{subfigure}
\usepackage{url}
\usepackage{printlen}
\usepackage{tikz}
\usepackage{pgfplots}
\usepackage{amsmath}
\usepackage{amsthm}
\usepackage{algpseudocode}

\usepackage{multirow,tabularx}
\newcolumntype{Y}{>{\centering\arraybackslash}X}

\newtheorem{theorem}{Theorem}
\newtheorem{problem}{Problem}

\newtheorem{ilp}{ILP}

\usepackage[utf8]{inputenc}

\usepackage[draft]{fixme}

\DeclareMathOperator{\intra}{intra}
\DeclareMathOperator{\inter}{inter}
\DeclareMathOperator{\ACK}{ACK}

\newcommand{\genname}{DCT${^2}$Gen}

\newcommand{\lt}{Layer~2}
\newcommand{\lf}{Layer~4}

\newcommand{\obstm}{\ensuremath{\mathrm{TM}^{(\mathrm{obs})}}}
\newcommand{\gentm}{\ensuremath{\mathrm{TM}^{(\mathrm{gen})}}}
\newcommand{\pltm}{\ensuremath{\mathrm{TM}^{(\mathrm{PL})}}}
\newcommand{\acktm}{\ensuremath{\mathrm{TM}^{(\mathrm{ACK})}}}

\newcommand{\bytes}[2]{\ensuremath{\mathrm{B}^{\mathrm{#1}}_{\mathrm{#2}}}}
\newcommand{\partners}[2]{\ensuremath{\mathrm{N}^{\mathrm{#1}}_{\mathrm{#2}}}}
\newcommand{\iat}[1] {\ensuremath{\mathrm{IAT}^{\mathrm{#1}}}}
\newcommand{\size}[1]{\ensuremath{\mathrm{S}  ^{\mathrm{#1}}}}

\uselengthunit{mm}

\begin{document}

\title{\genname{}: A Versatile TCP Traffic Generator for Data Centers}
\author{\IEEEauthorblockN{Philip Wette, Holger Karl}\\
\IEEEauthorblockA{University of Paderborn\\
33098 Paderborn, Germany\\
Email: \{philip.wette, holger.karl\}@uni-paderborn.de}
}

\maketitle

\begin{abstract}
	Only little is publicly known about traffic in non-educational data centers.
	Recent studies made some knowledge available, which gives us the opportunity to create more realistic traffic models for
	data center research.
	We used this knowledge to create the first publicly available traffic generator that produces realistic traffic between hosts in 
	data centers of arbitrary size.
	We characterize traffic by using six probability distribution functions and concentrate on the generation 
	of traffic on flow-level.
	The distribution functions are described as step functions, which makes our generator highly configurable 
	to generate traffic for different kinds of data centers.
	Moreover, in data centers,
	traffic between hosts in the same rack and hosts in different racks have different properties.
	We model this phenomenon, making our generated traffic very realistic.	
	We carefully evaluated our approach and conclude that it reproduces these characteristics with high accuracy.

	%The output of our traffic generator is a schedule of data transmissions describing where a flow of which size is initiated between which hosts.
	
\end{abstract}

\section{Introduction}
%\printlength{\columnwidth}

Traffic traces from data-center networks are very rare. This leads to problems when evaluating new networking ideas for data centers because
it is not possible to find proper input. 
We propose a method to generate realistic traffic for arbitrarily sized data centers.

Recent studies \cite{MSR-datacenters, datacentersInTheWild} investigated traffic patterns in today's data centers on flow level.
They gave a detailed statistical description for both the traffic matrices and the flows present on Layer~2 in data centers.
These studies were the first to give a detailed insight into the communication patterns of commercial data centers
and reported that different parts of the traffic matrix have different statistical properties. 
This is due to 
%the applications running in these data centers, which are aware of the organizational structure of servers in racks.
software that is tuned for running in a data center (like Hadoop \cite{hadoop}). The applications try to keep as much traffic in the same rack as possible
to achieve a higher throughput and lower latency. We call this property \emph{rack-awareness}.

We propose 
the Data Center TCP Traffic Generator (\genname{})
which takes a set of Layer~2 traffic descriptions and uses them to generate Layer~4 traffic for data centers.
When the generated Layer~4 traffic is transported using TCP, it results in Layer~2 traffic complying with the given descriptions.
With the Layer~4 traffic at hand, TCP dynamics can be included into the evaluation of novel networking ideas for data centers
with pre-described properties of Layer~2 traffic.
Our generator is highly realistic; e.g. it reflects rack-awareness of typical data center applications
which enables highly realistic evaluation of novel data center ideas.

\begin{figure}
	\centering
	\includegraphics[scale=1]{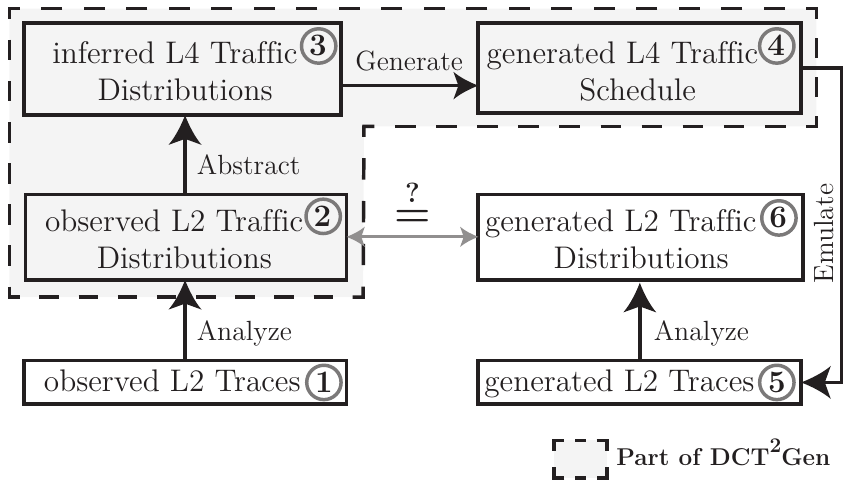}
	\caption{High-level overview of \genname{}s work flow.}
	\label{fig:workflow}
\end{figure}

The work flow required to generate artificial TCP traffic and to proof its validity is depicted in Figure~\ref{fig:workflow}.
First, Layer~2 traces from the targeted data-center are collected (1).
Then, these traces are analyzed to
obtain a set of probability distributions describing the traffic (2).
These distributions include the number of communication partners per host, the flow sizes, 
the sizes of traffic matrix entries, and others.
From the observed Layer~2 traffic distributions (2) we infer the underlying Layer~4 traffic distributions (3).
Using these \lf{} distributions, 
we generate a Layer~4 traffic schedule (4). 
This schedule describes for each host when to send how much \emph{payload} to which other host in the data center.
%All payloads are transported using TCP.
We claim that by executing our calculated schedule, the resulting traffic on Layer~2 (5) has the same stochastic properties as 
the original Layer~2 traffic traces (1).
To proof that, we are using network emulation to execute the computed \lf{} schedule. We capture the resulting traffic at Layer~2 (5) from the emulation and
analyze its statistical properties (6) to show that these are the same for both the original Layer~2 traces (1) and the generated traces (5).

To compute the Layer~4 traffic schedule, we do not even need to know the Layer~2 traces (1). 
It is sufficient to know the Layer~2 traffic distributions (2).
Even if for a data center it is possible to directly obtain \lf{} traffic distributions (3), \genname{} serves as a useful tool to generate
\lf{} schedules (4) from this data. Because even then, it is still necessary to generate traffic matrices from the data and create TCP flows complying with the
given distributions.

Finding a \lf{} traffic schedule (4) is a challenging task because of the bidirectional nature of TCP.
TCP is the most common Layer~4 protocol. For this work, we assume that all Layer~4 traffic is transported using TCP and that all TCP connections
are \emph{non-interactive} (i.e., payload is only transported into one direction).
In TCP, each flow transferring payload between a source $s$ and a destination $d$ also creates a flow of acknowledgments (ACKs) from $d$ to $s$.
The size of this ACK flow is roughly proportional to the size of transferred payload.
Thus, half of all flows in the schedule cannot be scheduled arbitrarily. The properties of these flows are depend on the other half of flows.
This poses a lot of interesting problems that we solved when creating our traffic generator.

\genname{} is open source and available for download from our website\footnote{\url{http://www.upb.de/cs/cn}}. 
%We supply all necessary inputs required to generate data center traffic which makes it very simple for others to use our traffic generator.
We supply all necessary inputs required to generate a traffic schedule complying with the distributions reported in \cite{MSR-datacenters, datacentersInTheWild}.
\genname{} can also be used to create traffic that has properties that differ from the ones used in this work.
In that case, solely the probability distributions (which are given as step functions and are part of the input) have to be adjusted accordingly.

The rest of this paper is structured as follows. In Section~\ref{sec:relatedWork} we give a short overview of the landscape of 
traffic generators. Section~\ref{sec:traffic} discusses traffic properties of data-center networks. These properties have to be replicated by 
our traffic generator whose architecture is presented in Section~\ref{sec:architecture}.
Section~\ref{sec:deconvolve} deals with one of the main challenges of this work:
A method is described to find the
distribution of the sizes of Layer~4 traffic matrix entries from the distribution of the sizes of Layer~2 traffic matrix entries.
%which are given as an input to \genname{}.
Section~\ref{sec:tm} describes the process of traffic matrix generation. Section~\ref{sec:flows} explains how to use these traffic matrices
to create a schedule of Layer~4 traffic. In Section~\ref{sec:evaluation} we evaluate our traffic generator and conclude this paper
in Section~\ref{sec:conclusion}.

\section{Related Work}
\label{sec:relatedWork}
Past research has created a large number of different traffic generators, all with different aims and techniques.
From our point of view, there are four key characteristics of available traffic generators:
\begin{itemize}
	\item Flow-level vs. packet-level
	\item Traffic on one link only vs. traffic on a whole network
	\item Automatic vs. manual configuration
	\item Topology awareness vs. non topology awareness
\end{itemize}
We give a short overview of each characteristic and afterwards use them to categorize existing traffic generators.

\subsection{Flow-level vs. packet-level generators}
There are traffic generators that output traffic on packet level, formatted due to 
certain communication protocols. The mix of these packets follows certain rules and probability distributions that are configurable beforehand.
However, these traffic generators \cite{itg, d-itg, rude} do not usually implement flows, i.e. packets that logically belong together and that
share certain properties like source and destination addresses. 
A traffic generator that is flow-aware \cite{swing,barakat2003modeling,conf/sigmetrics/SommersKB04} always generates packets organized in flows.
Flow generation is done such that the flows meet certain statistical properties.

\subsection{Traffic on one link only vs. traffic on a whole network}
The majority of existing traffic generators concentrates on generating traffic originating from one interface only.
For performance evaluation of whole network topologies it is required to know the packet stream that is created by each single device in the network.
As typically these streams are correlated, it is not sufficient to generate traffic for each interface separately but a traffic generator that 
creates correlated traffic for a whole network is required.

\subsection{Automatic vs. manual configuration}
Network traffic has various properties depending on the type of the network. 
To specify desired traffic properties, traffic generators can be parameterized by hand \cite{heegaard2000gensyn},
automatically by feeding \emph{traffic traces} from a real network whose traffic 
has to be mimicked \cite{weigle2006tmix, conf/sigmetrics/SommersKB04, Siska:2010:FTG:1815396.1815503}, or 
by a combination of both \cite{conf/sigmetrics/SommersKB04, Siska:2010:FTG:1815396.1815503}.
For the automatic case either algorithms are used to extract parameters from the given traffic or the given traffic
itself is part of the generated traffic. In that case it is often used as background traffic that is superimposed by 
special traffic that has properties based on the desired test case.

\subsection{Topology awareness vs. non topology awareness}
Traffic generators can be \emph{topology-aware}. 
In that case, the topology of the target network influences the traffic patterns produced by the generator.
In the context of data-center networks, the traffic matrices are typically dense in intra-rack areas and coarse in inter-rack areas.
Thus, a traffic generator for data center networks has to account for the placement of servers in racks.

\subsection{Existing Traffic Generators}
Harpoon \cite{conf/sigmetrics/SommersKB04} is an open source traffic generator that creates traffic at flow level.
It creates correlated traffic between multiple endpoints and automatically derives 
traffic properties from supplied packet traces. Harpoon is able to generate both TCP and UDP traffic. The general concept of Harpoon is a
hierarchical traffic model. 
Traffic between any pair of endpoints is exchanged in sessions where each session consists of multiple file transfers between
that pair of hosts. Sessions can either be TCP or UDP. Harpoon can be parametrized in terms of inter-arrival and holding times for
sessions, flow-sizes, and the ratio between UDP and TCP.
These parameters are automatically derived from supplied packet traces.
As Harpoon is not topology-aware it cannot be used to replicate the special properties of data-center traffic.

Ref.~\cite{Siska:2010:FTG:1815396.1815503} proposes a flow-level traffic generator for networks. It uses a learning algorithm that 
automatically extracts properties from 
packet traces. 
That work focuses on generation of traffic from different applications each with different communication patterns.
To this end, \emph{Traffic Dispersion Graphs} are used to model the communication structure of applications.
The generator reproduces these communication structures accurately but is less accurate in modeling the properties of flows.
In addition, this traffic generator does not capture any structural properties of the traffic matrix.
%They draw flow sizes from a Lognormal distribution which does not reflect the situation in data centers.

The Internet Traffic Generator \cite{itg} and its distributed variant D-ITG \cite{d-itg} focus on traffic generation on packet-level.
Both generate a packet stream that can be configured in terms of the inter-departure time and the packet size.
A similar traffic generator is presented in \cite{barakat2003modeling}. It generates traffic on flow level for a single internet backbone link.

Swing \cite{swing} is a closed-loop traffic generator that uses a very simple model for generating traffic on packet level.
Swing aims at reproducing the packet inter-arrival rate and its development over time. Packets are logically organized in flows.
However, Swing only generates a packet trace for a single link.

Up to now, there exists no traffic generator that computes a schedule of TCP payload transmissions 
that can be used to produce Layer~2 traffic with predescribed properties.
\genname{} is the first generator to compute such a schedule.

\section{Traffic Properties}
\label{sec:traffic}

To describe and generate traffic, \genname{} uses several stochastic traffic properties. Some of these properties are observed from L2 traces that are given as input, some of them are properties of inferred Layer~4 traffic. The traffic description hinges on how flows behave inside the network. 

% This section describes the six stochastic traffic properties of Layer~2 traffic that are used as input to our traffic generator (Figure~\ref{fig:workflow}, box~2).
% Each of these properties is a random variable, described by a Cumulative Distribution Function (CDF).
% Table~\ref{table:cdf-summary} summarizes the CDFs that are used by  \genname{}, partially as input, partially as computed distributions.
%We are using step functions because this makes it very simple to reconfigure the generator for other traffic scenarios.
% The goal of \genname{} is to produce a schedule of TCP transmissions that produces traffic on Layer~2 with the given properties.

% \fxnote{Ist der Absatz hier sinnvoll?}All inputs to \genname{} describe properties of the targeted traffic on Layer~2.
% There is no input describing any Layer~4 properties because in general it is not possible to retrieve this information from 
% a network without modifying the end-hosts or using deep packet inspection. Thus, all inputs required to \genname{} can be retrieved
% using common network monitoring techniques.

Throughout the paper, the term \emph{flow} describes a 
series of packets on Layer~2 between the same source and destination that logically belong together. We distinguish two types of flows. 
A \emph{payload flow} is a flow which transports payload from source to destination -- looking from Layer~4, it transports TCP data packets. Since we assume non-interactive TCP traffic, a payload flow does not include any acknowledgments. Acknowledgments are sent in separate \emph{ACK flows}, which only include TCP ACK segments but no data. In consequence, each TCP connection results in two flows. The structure of these flows is captured by \emph{traffic matrices} described in the following.

%The distribution of the size of flows at Layer~2 is called the \emph{flow-size distribution} (Section~\ref{sec:flow-sizes}). 
%The distribution of the size of payload flows is called \emph{payload-size distribution}
%and the distribution of the size of ACK flows is called \emph{ACK-size distribution} (both described in Section~\ref{sec:flow-inter-arrival}).

\subsection{Traffic Matrices}
\label{sec:traffic-matrices}

A traffic matrix (TM) describes the \emph{amount of data} in bytes (not number of flows) that is transferred between a set of end hosts in a fixed time interval, capturing the pattern of flows in such a time interval. The entry $(i,j)$ of a TM tells how much data is sent from server $i$ to server $j$ during that time.

We distinguish several types of traffic matrices. The primary one is the traffic matrix describing the \emph{observed,} actual traffic on Layer~2; we denote this matrix as \obstm. The next matrix corresponds to the \emph{generated} traffic on Layer~2 that is a result of \genname{} (compare Box~5 in Figure~\ref{fig:workflow}). Generated L2 traffic is described by the traffic matrix \gentm. 

\lt{} traffic is juxtaposed to \lf{} traffic. \lf{} traffic is obviously not observed but only generated. We have to distinguish between the \emph{payload} traffic matrix describing the actual data flows and the \emph{acknowledgement} traffic matrix for the flows containing only acknowledgement packets; they are called \pltm{} and \acktm, respectively. Since the payload flows from $i$ to $j$ give raise to the acknowledgement flows from $j$ to $i$, these two matrices are interrelated: 
$$ \acktm(i,j) = \beta \cdot \pltm(j,i)$$ \noindent for some value $\beta$ to be discussed in Section~\ref{sec:tm}. 

Moreover, a \lt{} traffic matrix is the sum of a \lf{} payload TM and a \lf{} ACK TM plus overhead; Section~\ref{sec:deconvolve} discusses the overheads involved here. 

		% We distinguish between the Layer~2 TM (\ethtm{}) and the Layer~4 TM (\tcptm{}).
		% The \ethtm{} describes all traffic transmitted at Layer~2. 
		% Thus, the entry $(i,j)$ of a \ethtm{} is the number of payload bytes sent from $i$ to $j$ \emph{plus}
		% the bytes required to acknowledge payload sent from $j$ to $i$.
		% The corresponding \tcptm{} contains the payload bytes only.
		% When the amount of payload specified in the \tcptm{} is transferred using TCP, the resulting traffic on Layer~2 is described by \ethtm{}.
		
In addition to the layer, traffic matrices also reflect the traffic structure inside a data center. 
Because of rack-aware applications, traffic inside a rack has different stochastic properties than that between racks -- we reflect these differences by separate stochastic distribution functions for the \emph{intra-rack} and the \emph{inter-rack} parts of a traffic matrix. % (an entry $(i,j)$ of a traffic matrix is called intra-rack if both $i$ and $j$ are located in the same rack)
                
For either part, we have to describe first the stochastic distribution of the number of nodes a given node $i$ talks to (either in its own or in any other rack). Second, for each node $j$  to which node $i$ talks, we need a stochastic distribution to describe the amount of bytes that is transferred from $i$ to $j$; the intra- and inter-rack cases will have different distribution functions. 

In summary, we need stochastic distributions separately for (a)  the cases of observed and generated \lt{} traffic or for payload and ACK traffic on \lf{}, (b) the distinction between intra- and inter-rack traffic, and (c) the description of number of communication partners vs.\ number of transferred bytes. This results in sixteen distribution functions so far. 

		% This behavior divides data-center TMs (both on \lt and on \lf) into two different areas,\fxnote{part? region?} each with different stochastic properties.

		% Non-zero TM entries are coarse in the inter-rack areas and dense in the intra-rack areas.
		% In addition, the distribution of non-zero intra-rack TM entry sizes differs significantly from the inter-rack sizes.
		% This is why we describe these areas independently.
		% The two interesting properties of a traffic matrix are the distribution of the number of non-zero entries per host and the 
		% distribution of the non-zero TM entry sizes.
		% We thus describe a traffic matrix by the following four  random variables:
		% \begin{itemize}
		% 	\item Amount of intra-rack traffic between host pairs 
		% 	\item Amount of inter-rack traffic between host pairs 
		% 	\item Number of intra-rack communication partners per host
		% 	\item Number of inter-rack communication partners per host
		% \end{itemize}
		% The first two variables describe the size of the different non-zero traffic matrix entries while the last two describe the number
		% of non-zero entries per row. 

	\subsection{Flow Sizes}
\label{sec:flow-sizes}
\begin{figure}
	\centering
	\includegraphics[scale=1]{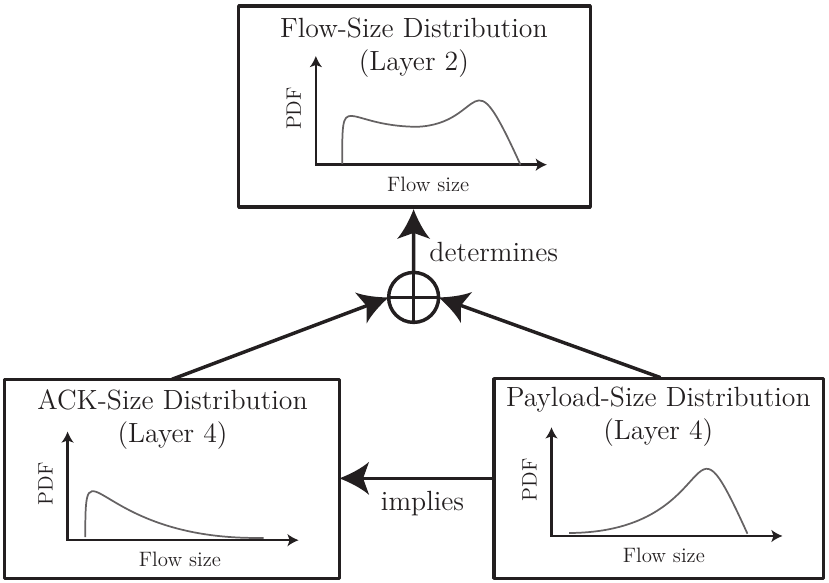}
	\caption{Relationship between the observable flow-size distribution function at Layer~2 (top), the distribution function of ACK sizes (bottom left) and
	causal payload sizes (bottom right).}
	\label{fig:flowsizeL2toL4}
\end{figure}

		The \emph{flow size} denotes the number of bytes transported by a flow including all protocol overhead.
%		The flow-size distribution is one of the most important properties of network traffic.
		An entry of a traffic matrix describes how much traffic is exchanged in total between a pair of nodes in a given time but it specifies neither
		number nor size of the individual flows transporting this traffic.
		The \emph{flow-size distribution} specifies how likely a flow of a certain size occurs.
		We distinguish between the flow-size distribution of payload flows \emph{and} ACK flows. 
		However, on Layer~2 we cannot tell which one is which.
		Flow sizes on Layer~2 are induced by the distribution of payload sizes exchanged by the servers on Layer~4 (TCP).
		Thus, we need a third distribution describing the distribution of flow sizes on \lt{}.
		In the process of traffic generation, \genname{} takes the Layer~2 flow-size distribution and computes the corresponding Layer~4 distribution 
		(Figure~\ref{fig:flowsizeL2toL4}).
%		This is the distribution of payload sizes which (when transported using TCP) creates
%		flows with sizes distributed according to the given Layer~2 flow-size distribution (Figure~\ref{fig:flowsizeL2toL4}).
		To be able to infer flow sizes at Layer~4 (from the flow sizes at \lt{}) we need to assume that all TCP sessions in the data center are non-interactive.
		For data centers running mostly Map-Reduce workload this assumption is true for most of the flows.
		
%		The distribution of flow sizes is specific to the mix of applications that runs on the network. 
%		Reference \cite{datacentersInTheWild} reports that this distribution is comparable among different data centers they inspected. 
%		However, in general the distribution is highly dependent on the type of application:
%		A data center that primarily hosts video content has a different flow size distribution than a data center
%		hosting social media services such as Facebook or Twitter where the single files that are transmitted are magnitudes smaller.
		
	\subsection{Flow Inter-Arrival Time}
\label{sec:flow-inter-arrival}
	The flow inter-arrival time distribution describes the time between two subsequent flows arriving at the network.
	Together with the flow-size distribution the flow inter-arrival time specifies the distribution of the total amount of traffic
	for a given time interval.
	This amount of traffic must match the total traffic specified by the TM.
	Otherwise, not enough (or too many) flows exist, which means it is not possible to use these flows to create a TM with the desired properties.

\subsection{Nomenclature}
\label{sec:nomenclature}

We shall indicate the distribution functions (\emph{not} the random variables) as follows: 
\begin{itemize}
\item $N$ represents the number of communication partners per node, \bytes{}{} represents the total bytes exchanged between a pair of nodes, \size{}{} represents the flow size and \iat{}{} represents flow inter-arrival times
\item the subscript specifies either the intra- or inter-rack case, if needed
\item the superscript specifies the case of observed or generated (on \lt) vs.\ payload or ACK  (on \lf) traffic
\end{itemize}

Table~\ref{table:cdf-summary} summarizes all the stochastic distribution functions that we use in the remainder of the paper. 

%\begin{table}
%\centering
%\caption{User-specified input to \genname{}. Each Cumulative Distribution Function is given as a step function and describes traffic at Layer~2.}
%\label{table:cdf-summary}
%\begin{tabular}{l|l}
%\hline 
%CDF~1 & Amount of intra-rack traffic between host pairs \\
%CDF~2 & Amount of Inter-rack traffic between host pairs \\
%CDF~3 & Number of Intra-rack communication partners per host \\
%CDF~4 & Number of Inter-rack communication partners per host \\
%CDF~5 & Flow sizes \\
%CDF~6 & Flow inter-arrival time \\
%\hline \\
%\end{tabular} 
%\end{table}

\begin{table*}
\centering
\caption{Overview of the distribution functions.}
\label{table:cdf-summary}
%\begin{tabularx}{\textwidth}{*{5}{Y|}}
\begin{tabularx}{\textwidth}{ll|Y||Y|Y|Y}
\multirow{2}{*}{Distributions} 	& 				& observed				& generated				& 	\multicolumn{2}{c}{inferred at \lf}		\\
								&				& \lt					& \lt					& PL				& ACK					\\
\hline
\multirow{2}{*}{Bytes}			& intra-rack		& \bytes{obs}{intra}		& \bytes{gen}{intra}		& \bytes{PL}{intra}	& \bytes{ACK}{intra}	\\
\cline{2-6}
								& inter-rack		& \bytes{obs}{inter}		& \bytes{gen}{inter} 	& \bytes{PL}{inter}	& \bytes{ACK}{inter} \\
\hline
\hline
\multirow{2}{*}{Number}			& intra-rack		& \partners{obs}{intra}	& \partners{gen}{intra}	& \partners{PL}{intra}& \partners{ACK}{intra} \\
\cline{2-6}
								& inter-rack		& \partners{obs}{inter}	& \partners{gen}{inter}	& \partners{PL}{inter}& \partners{ACK}{inter} \\
\hline
\hline
\multicolumn{2}{l|}{Flow Size}					& \size{obs}			& \size{gen}			&	\size{PL}		& \size{ACK}		\\
\hline
\multicolumn{2}{l|}{Flow Inter-Arrival time}		& \iat{obs}			& \iat{gen}			&	\iat{PL}		& \iat{ACK}		\\
\end{tabularx} 
\end{table*}

\section{Architecture of \genname{}}
\label{sec:architecture}

We generate a schedule of Layer~4 traffic
that specifies at which time how much \emph{payload} has to be transmitted from which source node to which destination node.
When transported using TCP, the generated \lt{} traffic on the network has the same properties as the observed \lt{} traffic.
To find this schedule, we are using the following approach.

First, a \pltm{} is generated.
To this end, we need to infer \bytes{PL}{inter} from \bytes{obs}{inter} and \bytes{PL}{intra} from \bytes{obs}{intra}.
From these distributions, along with \partners{obs}{inter} and \partners{obs}{intra}, a \pltm{} can be generated (for details, see Section~\ref{sec:tm}).
The next step is to assign flows to all non-zero \pltm{} entries.
For this task, we need to infer \size{PL} from \size{obs}.
%To this end, the \emph{payload}-size distribution function (Layer~4) has to be recovered from the flow-size distribution function (Layer~2). 
The former describes the distribution of
the sizes of payload flows that (together with the implied \size{ACK}) generates the given flow-size distribution on Layer~2
(Figure~\ref{fig:flowsizeL2toL4}).
Using \size{PL}, a set of payload flows is generated which are mapped to the non-zero \pltm{} entries in a subsequent step (Section~\ref{sec:flows}).

At the end of this process we know how many payload bytes to send from which node to which other node in TCP sessions such 
that it holds that: 
\bytes{gen}{intra} equals \bytes{obs}{intra}, 
\bytes{gen}{inter} equals \bytes{obs}{inter},
\partners{gen}{intra} equals \partners{obs}{intra},
\partners{gen}{inter} equals \partners{obs}{inter},
\size{gen} equals \size{obs},
\iat{gen} equals \iat{obs}
%\begin{itemize}
%	\item \bytes{gen}{intra} equals \bytes{obs}{intra}
%	\item \bytes{gen}{inter} equals \bytes{obs}{inter}
%	\item \partners{gen}{intra} equals \partners{obs}{intra}
%	\item \partners{gen}{inter} equals \partners{obs}{inter}
%	\item \size{gen} equals \size{obs}
%	\item \iat{gen} 		equals \iat{obs}
%\end{itemize}

\begin{figure}
	\includegraphics[scale=1.05]{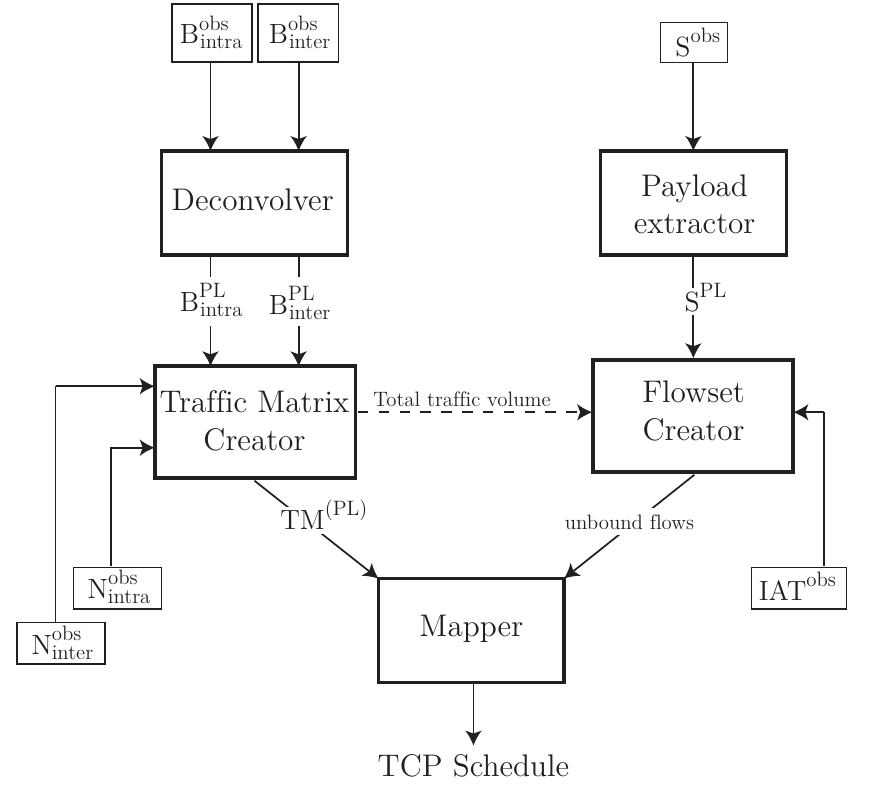}
	\caption{Architectural overview of \genname{}.}
	\label{fig:architecture}
\end{figure}

The modular design of our traffic generator can be seen in Figure~\ref{fig:architecture}. It consists of the five different modules
\emph{Deconvolver}, 
\emph{Payload Extractor},
\emph{Traffic Matrix Generator},
\emph{Flowset Creator}, and 
\emph{Mapper}.
%The required inputs to the traffic generator are listed in Table~\ref{table:cdf-summary}.
This section gives a short description of each single module. In the subsequent sections, complex modules 
(Deconvolver, Traffic Matrix Generator, Mapper) are explained in detail.

\subsection{Deconvolver}
The \emph{Deconvolver} takes \bytes{obs}{intra} and \bytes{obs}{inter} as inputs.
From these, it computes \bytes{PL}{intra} and \bytes{PL}{inter},
which enable us to generate \pltm{}.
As the name suggests, the Deconvolver is using a deconvolution technique which is explained in detail in Section~\ref{sec:deconvolve}.

\subsection{Payload Extractor}
We need to compute a set of payload flows that, together with the implied ACK flows, generate flows on \lt{} which comply with \size{obs}
(Figure~\ref{fig:flowsizeL2toL4}). 
Flows on \lt{} are the union of payload flows and ACK flows.
As we are only given \size{obs} we need to infer \size{PL}
(which itself implies a certain \size{ACK}).
\size{PL} is computed in the \emph{Payload Extractor}.

For the Payload Extractor to work, we need to
assume that the ratio of ACK packets to payload packets in TCP is fixed at a value $r$. 
We substantiate this assumption in Section~\ref{sec:payloadtoackratio}.
Once a concrete value $r$ is known, the Payload Extractor transforms the \size{obs} into \size{PL}.

Figure~\ref{algo:flowsizetoPL} shows a simple algorithm to infer \size{PL} from \size{obs}.
Let $\Pr^{\mathrm{obs}}(x)$ be the probability (according to \size{obs}) that the size of a flow is $x$
and $\Pr^{\mathrm{PL}}(x)$ the probability (according to \size{PL}) that the size of a payload flow is $x$.
$\ACK\left(x\right) = x \cdot r \cdot 66$ is the size of an ACK flow acknowledging
the receipt of $x$ payload bytes. $66$ is multiplied because in TCP an ACK packet has a size of 66 bytes.
To convert \size{obs} into \size{PL},
the algorithm iterates over all flow sizes in descending order
and removes the corresponding ACK flow from \size{PL}.
This works because it always holds that the ACK-flow size is smaller than or equal to the corresponding payload-flow size.

\begin{figure}
\begin{algorithmic}[1]
	\State \textbf{Algorithm} \textsc{InferPayloadSize}$\left(\,Pr^{\mathrm{obs}}(\cdot)\,\right)$:
	\State $\Pr^{\mathrm{PL}}(\cdot) \gets Pr^{\mathrm{obs}}(\cdot)$
	\For{\textbf{each} flow size $x$ in decreasing size}
		\State $\Pr^{\mathrm{PL}}\left(\ACK\left(x\right)\right) \gets \Pr^{\mathrm{PL}}\left(\ACK\left(x\right)\right) - \Pr^{\mathrm{PL}}\left(x\right)$
	\EndFor
	\State $\Pr^{\mathrm{PL}}(\cdot) \gets \Pr^{\mathrm{PL}}(\cdot) / \sum_{x} \Pr^{\mathrm{PL}}(x)$ \Comment{normalize $\Pr^{\mathrm{PL}}(\cdot)$}
	\State \textbf{return} $\Pr^{\mathrm{PL}}(\cdot)$
\end{algorithmic}
	\caption{Algorithm to transform \size{obs} to \size{PL}.}
	\label{algo:flowsizetoPL}
\end{figure}

\subsection{Traffic Matrix Generator}
From the outputs of the Deconvolver (\bytes{PL}{intra} and \bytes{PL}{inter}) together with \partners{obs}{intra} and \partners{obs}{inter},
the \emph{Traffic Matrix Generator} creates a \pltm{}.
This \pltm{} specifies payloads such that, when exchanged using TCP, this results in a \gentm{}
having the same statistical properties as \obstm{}.
Matrix generation is explained in detail in Section~\ref{sec:tm}.
After the \pltm{} has been calculated, the payloads exchanged between any pair of hosts are divided into single payload flows.

\subsection{Flowset Creator}
Flows are generated by the \emph{Flowset Creator}.
The Flowset Creator gets \size{PL} from the Payload Extractor, \iat{obs}, and 
a target traffic volume (which is the sum over all entries of the \pltm{} generated in the previous step). 
It outputs a set of flows whose flow sizes sum up to the target traffic volume.
%To this end, the Flowset Creator creates payload flows complying with \iat{PL} and \size{PL}. 

To this end, the Flowset Creator creates payload flows with sizes distributed according to \size{PL}.
When the payload flows are transferred over the network, the generated \lt{} flows have to comply with \iat{obs}.
For this task, we need to infer \iat{PL} from \iat{obs} such that \iat{PL} and \iat{ACK} result in \iat{obs}.
However, the resolution of the data provided by \cite{MSR-datacenters} for \iat{obs} is so low that we could not
draw any conclusions on \iat{PL}. This is why we use \iat{obs} as an approximation for \iat{PL}; this is generally not a 
proper approximation, however, we do not have any better data at hand.

The generated flows only add up to the target traffic volume if \iat{PL} and \size{PL}
are chosen such that the sum of all generated flow sizes matches the traffic volume of the generated \pltm{}.
If this is not the case, we scale the inter-arrival times by a linear factor to generate more or less flows depending on the situation.

%If the generated flows do not sum up to the desired traffic volume, the 
%inter flow arrival time distribution is shrinked or stretched and the process is repeated accordingly.

\subsection{Mapper}
In the last step of our traffic generator the flows generated by the Flowset Creator are mapped to the source-destination pairs specified by the
\pltm{} as computed by the Traffic Matrix Creator. 
This mapping is done by the \emph{Mapper} which uses a newly developed assignment strategy. The Mapper 
and our mapping strategy are explained in detail in Section~\ref{sec:flows}.

\section{Deconvolving Traffic Matrix Entries}
\label{sec:deconvolve}
\subsection{Problem description}
\label{sec:deconvolve_desc}
	The outcome of the traffic generation process is a schedule of payload transmissions specifying when which amount of payload is sent from one machine
	to another. 
	In TCP, whenever a certain amount of payload is transferred over the network, this payload flow causes a second flow called ACK flow.
	The ACK flow acknowledges the correct reception of the payload flow
	but does not transmit any payload itself.\footnote{Data exchange between two hosts over different TCP connections is of course supported by \genname{}.} 
	However, it adds traffic to the network. 
	The traffic seen on \lt{} is the sum of the payload flows and the ACK flows.
	We only have information from the observed \obstm{} but we want to build the inferred \pltm{}. 
	To this end, we need to compute \bytes{PL}{intra} from \bytes{obs}{intra} and \bytes{PL}{inter} from \bytes{obs}{inter}
	Or, put in other words, we need to infer the \lf{} distributions of non-zero TM entry sizes from the corresponding \lt{} distributions.
	This section shows how to do that.
		
	The individual non-zero traffic matrix entry sizes of a \obstm{} can be expressed as random variables $Z = X+Y$ where $X$ and $Y$
	specify the amount of outgoing payload Bytes ($X$) and the amount of outgoing ACK Bytes ($Y$).
	The distribution of $Z$ is given as \bytes{obs}{inter} resp. \bytes{obs}{intra} and it is the \emph{linear convolution} of the distributions of $X$ and $Y$,
	which we both do not know.
	
	When assuming that the ratio between payload packets and ACK packets is a constant $r$ and by ignoring the facts that a) the TCP protocol
	adds overhead to each single packet and b) TCP uses additional messages for establishing and terminating sessions (TCP handshake), we can write $Z$ as
	\[Z = X + \beta Y\]
	where $\beta = \frac{|\mathrm{ACK}|}{|\mathrm{PAY}|} \cdot r$ is the size of an ACK packet divided by the size of a payload packet times $r$,
	recasting the ratio of packets $r$ into a ratio of bytes $\beta$.
	We treat $X$ and $Y$ as independent and identically distributed (iid) random variables
	although $X$ and $Y$ might be correlated.
	However, we assume that this correlation is very low for the kind of software that runs in a data center.
	For instance, when using Hadoop there is no reason to assume that the amount of payload 
	sent from server $i$ to server $j$ depends on the amount of payload sent from $j$ to $i$ (which determines how many ACKs are sent from $i$ to $j$).

	RFC~1122 \cite{rfc1122} states that for TCP is is not required to acknowledge the correct receipt of every single payload packet. 
	Instead, one ACK can acknowledge multiple payload packets at once.
	This is called \emph{delayed ACK}. However, the acknowledgement of payload must not be arbitrarily delayed.
	According to RFC~1122, the delay must be less than 0.5 seconds and 
	there has to be at least one ACK packet for every second payload packet.
	%From this specification it follows that for a backlogged TCP transmission that transfers sufficiently many bytes, $r$ converges 
	%to 2. 
	Unfortunately, TCP implementations in modern Operating Systems do not follow this specification strictly.
	In experiments with Linux Kernel 3.11, e.g., the number of outstanding unacknowledged payload packets ranged up to 16 for a
	backlogged 1\,GByte flow over a 1\,Gbit/s link.
	
	In the following, we show that for the TCP connections transferring most Bytes in a data center it holds that on average 
	the ratio between payload packets and ACK packets $r \approx 2.5$.
	
	%The problem solved in this section is how to infer the distribution of $X$ when the distribution of $Z$ and $r$ are known.
	%We afterwards apply the solution to infer
	%\bytes{PL}{intra} from \bytes{obs}{intra} 
	%and
	%\bytes{PL}{inter} from \bytes{obs}{inter}.
%	Thus, for the inter-rack case, $Z$ is distributed according to the 
%	distribution of the amount of inter-rack traffic between host pairs and $X$ according to the
%	distribution of the amount of inter-rack payload traffic between host pairs.
%	The intra-rack case is accordingly.
	
\subsection{Estimating Payload to ACK Ratio}
\label{sec:payloadtoackratio}

	We now show that $r$ is nearly constant in our data-center scenario.
	Clearly, the value of $r$ depends on a) the available link speeds, b) the TCP implementation, and c) the distribution of flow sizes.
	We want to calculate $r$ for payload-flow sizes distributed according to \size{PL}.
	To determine $r$, we have to compute a \pltm{} and divide its non-zero entries into payload flows. 
	Then, this traffic can be emulated using a network emulator and
	the resulting $r$ value can be observed. However, we know neither \pltm{} nor \size{PL}. 
	To compute both we need to know $r$ first, which means we are stuck in a vicious circle.
	
	The only pragmatic way of breaking the circle is to use \obstm{} as an approximate for \pltm{},
	divide the non-zero entries into payload flows distributed according to \size{obs} and emulate this traffic to determine $r$.
	This of course has a negative influence on the accuracy of the estimated $r$ value.
	However, since a) there is no reason why the traffic matrix should have a large effect on $r$ (as long as the payload sizes keep the same),
	and b) \size{obs} and \size{PL} are not too way off, the introduced error will be acceptable.

	To estimate $r$ we calculate a \obstm{}, 
	generate TCP traffic with payload sizes distributed according to \size{obs},
	and emulate this traffic using a network emulator.
	In a subsequent step we analyze the generated ACK packets to approximate $r$.

	We generated 60\,s of TCP traffic for a data center
	consisting of 1440 servers organized in 72 racks of 20 servers each, interconnected in a Clos-like topology (for details, see Section~\ref{sec:evaluation}).
	We emulate this data center with the network emulator MaxiNet \cite{wette14b}.
    \mbox{MaxiNet} distributes a network emulation across several physical machines
	to be able to emulate large networks under high traffic. We use a time dilation factor of 300 
	on a cluster of 12 servers equipped with Intel Xeon E5506 CPUs running at 2.16 Ghz.

\begin{figure}
	\centering
	\includegraphics[scale=0.5]{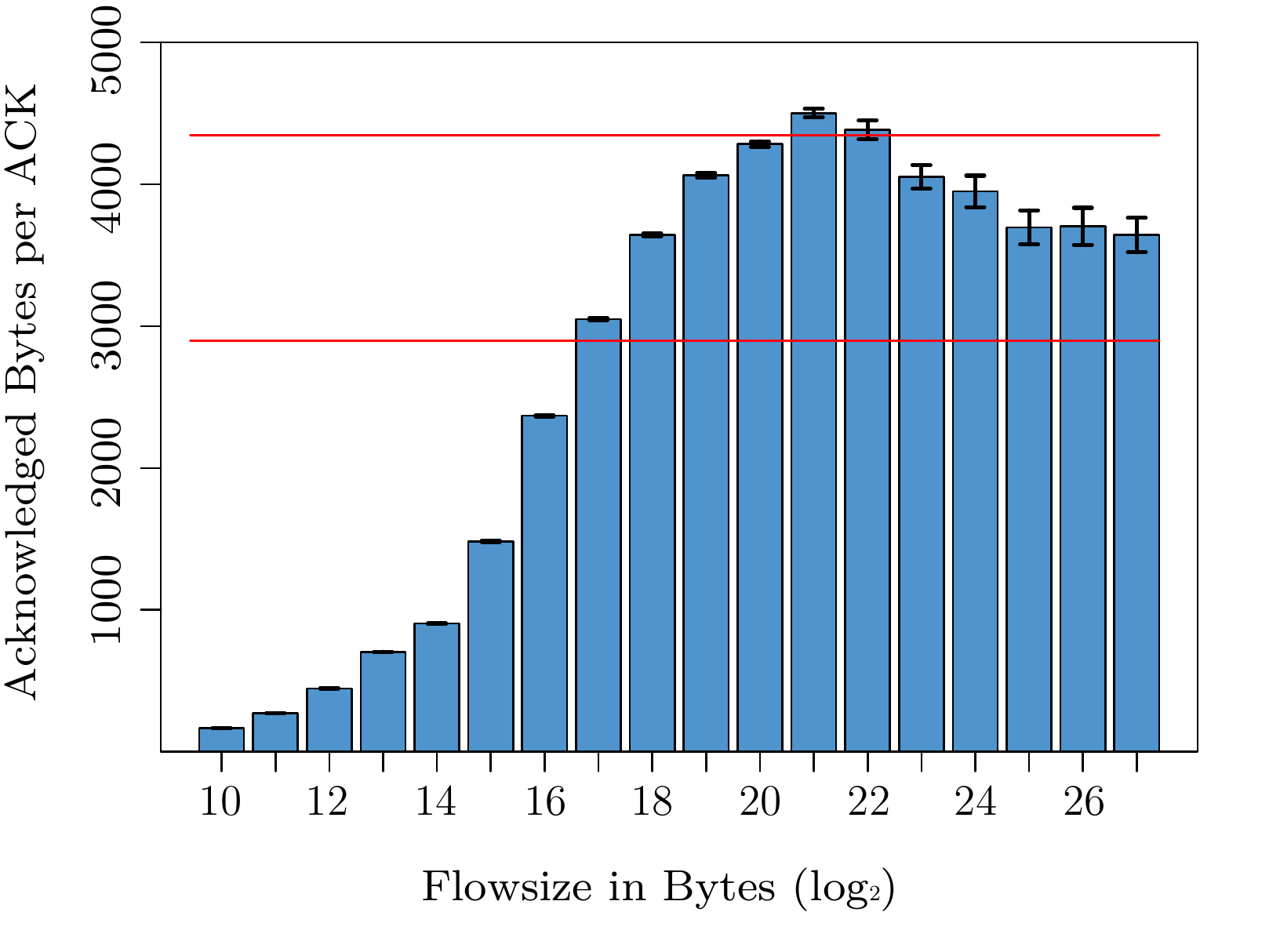}
	\caption{Number of acknowledged payload Bytes per ACK packet plotted over different flow sizes. Error bars show confidence intervals of level 0.95.}
	\label{fig:payloadToAck}
	%figure kann mittels makeBarPlot in ackAnalyzor.R generiert werden.
\end{figure}

	On both emulated core switches we used \texttt{tcpdump} to write a trace of all packets passing the first interface.\footnote{For performance reasons, it was not possible to create traces at all switches or interfaces. So we decided to use the core switches to be able to see traffic from all parts of the network.}
	In a subsequent step we analyzed all ACK flows in the trace to determine the ratio between the transferred payload and the number of ACK packets.
	Figure~\ref{fig:payloadToAck} plots the number of Bytes acknowledged by each ACK packet against the size of the 429,491 identified payload flows.
	The two horizontal lines mark 2896\,Bytes and 4344\,Bytes, which is one ACK packet for every second resp. every third payload packet (we used an MTU of 1500 which means the MSS was 1448).
	It can be seen that for each flow larger than $2^{16} \approx 65$\,KB the ratio between payload packets and ACKs is between 2 and 3. For larger flows
	the ratio stabilizes at 2.5.
	
	Note that the observable TCP dynamics depend on various environment characteristics.
	TCP always adapts to the current network situation by delaying ACKs and by enlarging or decreasing the TCP window size resulting in different data rates
	for the single flows.
%	The latter is mostly dependent on the bandwidth delay product and the level of congestion in the network.
%	But these dynamics can also change by altering the used traffic mix in the network or by changing the MTU or MSS.
	Thus, the ratio between payload packets and ACKs we found for our scenario can differ from the ratio in other network setups.
	In that case a different $r$ value has to be given to \genname{}.
	
%	Note that due to the TCP dynamics this ratio is dependent on the available bandwidth, the latency, the used TCP implementation, the
%	Maximum Segment Size (MSS), and the distribution of the flow sizes.
%	Thus, $r$ has to be adjusted for every single network setup.
	%However, we found out that for time dilation factors between 200 and 900 in our scenario the ratio between ack and payload packets stays fixed.

\subsection{Deconvolving TCP Traffic}
	Once we know $r$, which tells how much overhead is created by the ACK packets, we can easily calculate $\beta$ as it only depends on the MTU.
	We are now going to show how to extract the distribution of $X$ from
	\[Z = X + \beta Y\]
	when $\beta$ and the distribution of $Z$ are known and $X$ and $Y$ are independent and identically distributed (iid).
	This result can then be used to to infer both
	\bytes{PL}{intra} from \bytes{obs}{intra} 
	and
	\bytes{PL}{inter} from \bytes{obs}{inter}.
	
	Let $\operatorname{f}(t)$ denote the characteristic function of $X$ which we want to calculate.
	Since the characteristic function of the sum of two independent random variables is the product of both their characteristic functions,
	we can write the characteristic function $\operatorname{g}(t)$ of $Z / \beta$ as
	\[\operatorname{g}(t) = \operatorname{f}(t) \operatorname{f}(\gamma t)\]
	where $\beta = \frac{|\mathrm{ACK}|}{|\mathrm{PAY}|} \cdot r$ and $0 < \gamma = \frac{1}{\beta} < 1$.
%	As the CDF of $Z$ is one of the inputs to our traffic generator and we are able to calculate $\beta$, 
%	we can calculate the empirical characteristic function $g(t)$ of $Z / \beta$.
	Using the results of \cite{arXiv:math/0306237} we can write
	\[ \operatorname{f}(t) = \prod_{k=0}^{\infty} \frac{\operatorname{g}(\gamma^{2k}t)}{\operatorname{g}(\gamma^{2k+1}t)} ~~.\]
	Evaluating $\operatorname{f}(t)$ on each point from the range of $g(\cdot)$ yields an approximation of the characteristic function of $X$.
	From the characteristic function, the density can be calculated by an inverse Fourier transformation.

\subsection{Results}
	To ascertain that the deconvolution yields reasonable results, we now show the results of the deconvolution of
	\bytes{obs}{inter} (as reported in \cite{MSR-datacenters}).
	We set $r$ to 2.5, thus $\beta = \frac{1}{2.5} \cdot \frac{66}{\mathrm{MSS}}$
	and compute the deconvolution to retrieve \bytes{PL}{inter}.
	
	Then, we compute the implied \bytes{ACK}{inter}
	based on \bytes{PL}{inter} (Figure~\ref{fig:flowsizeL2toL4}).
	The function $\ACK(p)$ is used to compute the size (in bytes) of an ACK flow corresponding to a payload flow of size $p$:
	\[ \ACK(p) = 66 \cdot \left\lceil \frac{p}{\mathrm{MSS} \cdot r} \right\rceil \]
	where in our setup MSS was set to 1448. 66 is multiplied because in TCP an ACK packet has a length of 66\,bytes.
	We calculate the resulting \bytes{gen}{inter} as the convolution of \bytes{PL}{inter} and \bytes{ACK}{inter} and
	compare it to \bytes{obs}{inter} to ascertain that the deconvolution was successful.
	%Note that a simple multiplication with $\beta$ would be wrong here since this would result in non-integral numbers of ACK packets.
	
\begin{figure}
	\centering
	\includegraphics[scale=0.63]{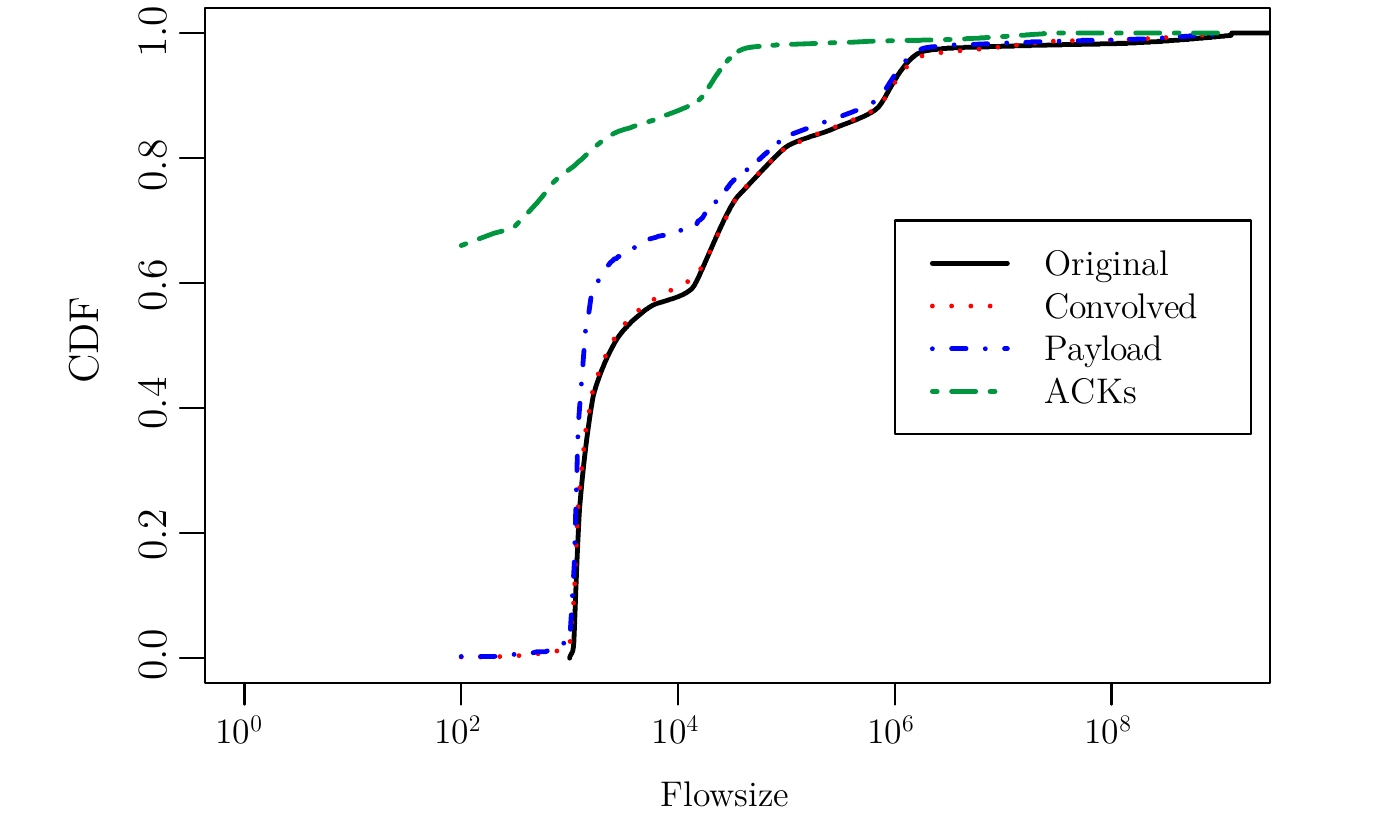}
	\caption{Result of the deconvolution operation. The solid line shows \bytes{obs}{inter} which is decomposed into 
	\bytes{PL}{inter} and the \bytes{ACK}{inter}.
	The dotted line depicts the convolution of \bytes{PL}{inter} and \bytes{ACK}{inter}.}
	\label{fig:convolution}
	%figure mittels smalltestoutOfRackFileSize aus extractPayloadDistribution.R erzeugt.
\end{figure}

	Figure~\ref{fig:convolution} shows the result of the deconvolution operation.
	It depicts the following four CDFs: 
	a) \bytes{obs}{inter} as the original CDF,
	b) \bytes{PL}{inter} as the inferred payload-size distribution, 
	c) \bytes{ACK}{inter} as the implied ACK-size distribution, and
	d) \bytes{gen}{inter} as the convolution of both \bytes{PL}{inter} and \bytes{ACK}{inter}.
	One can see that the CDFs of the original and the derived Layer~2 distribution are almost identical which shows that the deconvolution
	was successful.

	One should note that the deconvolution only yields an approximation of \bytes{PL}{intra} and might be noisy.
	This is primarily because the input distribution (\bytes{obs}{inter} in this case) is only an empirical approximation of a real distribution.
	In our case, we extracted \bytes{obs}{inter} from a figure in \cite{MSR-datacenters} which does not have a good resolution.
	The resulting noise is even more amplified by the transformation from the characteristic function to a 
	probability density function.
	We thus performed some manual filtering on the density function to retrieve the function depicted in Figure~\ref{fig:convolution}.
	This filtering basically removed negative values, smoothed the function, and scaled it to sum up to 1.

\section{Generating Traffic Matrices}		
\label{sec:tm}
		We generate a \pltm{} that specifies the amount of payload exchanged between server pairs within a fixed period.
		When this payload is transported using TCP, this generates a traffic matrix \gentm{} on \lt{}. For this TM it holds that:
		\begin{itemize}
			\item \bytes{gen}{intra} equals \bytes{obs}{intra}
			\item \bytes{gen}{inter} equals \bytes{obs}{inter}
			\item \partners{gen}{intra} equals \partners{obs}{intra}
			\item \partners{gen}{inter} equals \partners{obs}{inter}
		\end{itemize}
		
		To create a \pltm{}, we first determine each node's number of inter- and intra-rack communication partners by computing a random variable from 
		the corresponding distributions.	Then, we use the numbers as node degrees and look for such a 
		undirected simple graph\footnote{A simple graph is an undirected graph where no node has an edge to itself and no more than one edge between the same pair of nodes exists.} $G$.
		Finding a graph with a given inter- and intra-rack node degree 
		is the \emph{k-Partite Degree Sequence Problem} which is a variant
		of the intensively studied \emph{Degree Sequence Problem}. We give an Integer Linear Program (ILP) to solve the 
		k-Partite Degree Sequence Problem and study its run-time behavior.
		
		In a subsequent step we transform the adjacency matrix of $G$ into a traffic matrix by computing a random variable for the traffic volume 
		for each edge using \bytes{PL}{inter} resp. \bytes{PL}{intra}.

	\subsection{Problem Formalization}
		The problem of creating traffic matrices for $n$ nodes with given intra- and inter-rack node degrees 
		can be formalized as follows:
		The inter-rack node degree of a node is defined as the number of edges to nodes in different racks
		whereas the intra-rack node degree of a node is defined as the number of edges to nodes in the same rack.
		Let $V = \lbrace v_1, v_2, ..., v_n \rbrace$ be a set of vertices organized in racks of size $m$ where
		$v_{i \cdot m}, v_{i \cdot m +1}, ..., v_{(i+1) \cdot m -1}$ 
		$\forall$ $0 \leq i < \lceil n/m \rceil$		
		are located in the same rack $i$.
		Let $D^{\mathrm{int}} = (d^{\mathrm{int}}_1, d^{\mathrm{int}}_2, ..., d^{\mathrm{int}}_n)$ be the desired intra-rack node degrees and
		$D^{\mathrm{ext}} = (d^{\mathrm{ext}}_1, d^{\mathrm{ext}}_2, ..., d^{\mathrm{ext}}_n)$ the desired inter-rack node degrees for all $n$ nodes.
		We are looking for an undirected simple graph $G = (V,E)$ where the node degrees follow the intra- and inter-rack 
		degrees given by $D^{\mathrm{int}}$ and $D^{\mathrm{ext}}$.
	\subsection{The Degree Sequence Problem}
		\label{sec:GraphsWithDegs}
		Let $G = (V,E)$ be a simple graph on $n$ vertices.
		We call the decreasing order of the node degrees of $V$ the \emph{degree sequence} of $G$.
		\begin{problem}[The Degree Sequence Problem]
		\label{prob:DegSeqProb}
		Let $V = (v_1, v_2, ...,v_n)$ be a set of nodes and $D = (d_1, d_2, ..., d_n)$, $d_i \geq d_{i+1}$ $\forall$ $0 < i < n$, the desired node degrees.
		Find a simple 
		graph
		$G = (V,E)$, $E \subseteq V \times V$, where the degree sequence of $G$ is equal to $D$.
		If such a graph exists, $D$ is called \emph{realizable}.
		\end{problem}
		
		Problem~\ref{prob:DegSeqProb} is extensively studied \cite{hakimi62, havel55}.
		The main results on the Degree Sequence Problem for simple graphs are Theorem~\ref{theo:1} and Theorem~\ref{theo:2}.
		
		\begin{theorem}[Erdős–Gallai]
			$D$ is a realizable degree sequence for a simple graph on $n$ nodes if and only if
			\begin{enumerate}
				\item The sum of all desired node degrees is even
				\item $\sum_{i=1}^k d_i \leq k(k-1) + \sum_{i=k+1}^n \min(d_i, k)$ $\forall$ $1 \leq k \leq n$
			\end{enumerate}
			\label{theo:1}
		\end{theorem}

		\begin{theorem}
		$D = (d_1, d_2, ..., d_n)$ is realizable as a simple graph if and only if $D^\prime = (d_2-1, d_3-1, ..., d_{d_1+1}-1, d_{d_1+2}, d_{d_1+3}, ..., d_n)$
		is realizable as a simple graph.\\
			\label{theo:2}
		\end{theorem}

		From Theorem~\ref{theo:2} the iterative algorithm shown in Figure~\ref{algo:1} can be deduced to create a simple graph with a given node degree.
		The algorithm creates a graph for which it holds that $\deg(v_i) = d_i$ (if such a graph exists).

%		\begin{algo}		~
%			\begin{enumerate}
%				\item $G = (V,E)$, $V = \left\lbrace 1, 2, ..., n \right\rbrace $, $E = \left\lbrace \right\rbrace $
%				\item Let the initial residual node degree of node $v_i$ be $d_i$.
%				\item Let $U = (u_1, u_2, ..., u_n)$ be the list of vertices decreasing in the order of their residual node degree.
%				\item Create edges between $u_1$ and the next $d_1$ nodes in $U$. If no $d_1$ nodes exists with residual node degree $> 0$, abort.
%				\item Update $D$ as stated by Theorem~\ref{theo:2}.
%				\item If $U$ is not empty, goto 3.
%				\item Return $G = (V,E)$.
%			\end{enumerate}
%			\label{algo:1}
%		\end{algo}

\begin{figure}
\begin{algorithmic}[1]
	\State \textbf{Algorithm} \textsc{ConstructGraph}$\left(\, D = (d_1, ...,d_n) \,\right)$:
	\State $G = (V,E)$, $V = \left\lbrace 1, 2, ..., n \right\rbrace $, $E = \left\lbrace \right\rbrace $
	\State Let the initial residual node degree of node $v_i$ be $d_i$.
	\State Let $U = (u_1, u_2, ..., u_n)$ be the list of vertices decreasing in the order of their residual node degree.
	\State Create edges between $u_1$ and the next $d_1$ nodes in $U$.
	\If{no $d_1$ nodes exists with residual node degree $> 0$}
		\State \textbf{return} Error
	\EndIf
	\State Update $D$ and corresponding $U$ as stated by Theorem~\ref{theo:2}.
	\State If $U$ is not empty, goto 4.
	\State \textbf{return} $\Pr^\prime(\cdot)$
\end{algorithmic}
	\caption{Polynomial time algorithm to create a graph with a given node degree.}
	\label{algo:1}
\end{figure}

	\subsection{The k-Partite Degree Sequence Problem}

		Creating inter-rack edges is different from Problem~\ref{prob:DegSeqProb} because here there exist sets of nodes
		between which no edges are permitted.
		These sets are the sets of nodes located in the same rack.
		This leads us to
		Problem~\ref{problem:equalKpartiteDegreeSequence}, called the \emph{Degree Sequence Problem on $k$-Partite Graphs}.
%		\emph{k-Partite Degree Sequence Problem}.
%		\begin{problem}[The $k$-Partite Degree Sequence Problem]
%			Given $k$ degree sequences $D_1, D_2, ..., D_k$, find
%			an undirected $k$-partite Graph $G$ where each partition $i$ consists of $\mid D_i \mid$ nodes and for each node $v$ in partition
%			$i$ it holds that $\deg(v) = D_{i_v}$.
%			We call $D_1, D_2, ..., D_k$ \emph{valid} if such a graph exists.
%			\label{problem:kpartiteDegreeSequence}
%		\end{problem}
%		
%		In fact in our model all racks are of the same size $m$. So it suffices to solve the		
%		following more constrained version of the 
%		Degree Sequence Problem on Equally Sized $k$-Partite Graphs:
		\begin{problem}[Degree Sequence Problem on $k$-Partite Graphs]
			Given $k$ degree sequences $D_1, D_2, ..., D_k$, find
			an undirected $k$-partite Graph $G$ where each partition $i$ consists of $|D_i|$ nodes and for each node $v$ in partition
			$i$ it holds that $\deg(v) = D_{i_v}$.
			We call $D_1, D_2, ..., D_k$ \emph{realizable} if such a graph exists.
			\label{problem:equalKpartiteDegreeSequence}
		\end{problem}
%		Note that we can transform every instance of Problem~\ref{problem:kpartiteDegreeSequence} to an instance of 
%		Problem~\ref{problem:equalKpartiteDegreeSequence}  and vice versa by adding/removing nodes with zero degree.
%		
%			Given an intra rack degree sequence $D^\star = (d^\star_1, d^\star_2, ..., d^\star_n)$ and an integer $m \leq n$
%			where $n = k \cdot m$ for some $k \in \mathcal{N}^+$, find
%			an undirected $k$-partite Graph $G^\star$ which has node degrees $D^\star$ and where each partition consists of exactly $m$ nodes.
%			We call $D^\star$ \emph{valid} if such a graph exists.
%			\label{problem:kpartiteDegreeSequence}
%		\end{problem}

		%We will now show that both problems are $\mathcal{NP}$-Hard by a reduction from ??.
		
		Problem~\ref{problem:equalKpartiteDegreeSequence} is a special case of the 
		\emph{Restricted Degree Sequence Problem} \cite{Erdoes:arXiv1301.7523} in which arbitrary edges are forbidden to use.
		The best known algorithm to solve this problem requires to find a perfect matching on a simple graph of $\Omega(n^2)$ nodes.
		This makes this approach inapplicable to our problem: It already took more than 36 minutes on an Intel i7 2.2 Ghz processor to 
		calculate a graph for seven racks (each consisting of 20 servers) using the Boost graph library.
		
		Reference~\cite{milena02} presents the \emph{Degree Sequence Problem with Associated Costs} where the goal is to find
		a minimum cost realization of a given degree sequence. 
		It is possible to model Problem~\ref{problem:equalKpartiteDegreeSequence} when setting all costs for intra-rack edges to infinity.
		However, the running time to solve the problem is also dominated by finding a perfect matching on a graph with 
		$\Omega(n^2)$ nodes. We thus model our problem as an ILP with $n$ constraints, which is faster to solve for the problem instance sizes
		in this context.
		
		ILP~\ref{ilp:inter-rack} models Problem~\ref{problem:equalKpartiteDegreeSequence} where $D^{\mathrm{ext}} = D_1 \cup D_2 \cup ... \cup D_k$.
		In case that $D^{\mathrm{ext}}$ is realizable, ILP~\ref{ilp:inter-rack} computes a graph $G^{ext}$ with degree sequence $D^{\mathrm{ext}}$. 
		If $D^{\mathrm{ext}}$ is not valid it will compute the Graph $G^{ext}$ which has the highest possible edge count
		under the condition that no node has a higher degree than specified by $D^{\mathrm{ext}}$.
		
		\begin{ilp}{Constructing an Inter-Rack Graph}
			\begin{flalign*}
				\text{maximize} & \sum_{0 <i < n} \sum_{0 < j < i} b_{i,j} & & b_{i,j} \in \{0,1\} \\	
				\text{w.r.t.} & & \\
				& \sum_{j \in \inter(i)} b_{\max(i,j),\min(i,j)} \leq d^{\mathrm{ext}}_i & & \forall 1 \leq i \leq n
			\end{flalign*}
			\label{ilp:inter-rack}
		\end{ilp}
		
		In ILP~\ref{ilp:inter-rack}, $b_{i,j}$ equals 1 if an undirected edge exists between nodes $i$ and $j$.
		Note that ILP~\ref{ilp:inter-rack} only models the lower triangular matrix of the adjacency matrix of $G^{ext}$ because $G^{ext}$ is undirectional.
		$\inter(i)$ describes the set of nodes that are not in the same rack as node $i$ and is defined as
		\[ \inter(i) = \left\lbrace j \mid \left\lfloor \frac{j}{m} \right\rfloor \neq \left\lfloor \frac{i}{m} \right\rfloor ~\forall~ 0 < j \leq n \right\rbrace. \]
		
\begin{figure}
	\centering
	\includegraphics[scale=0.41]{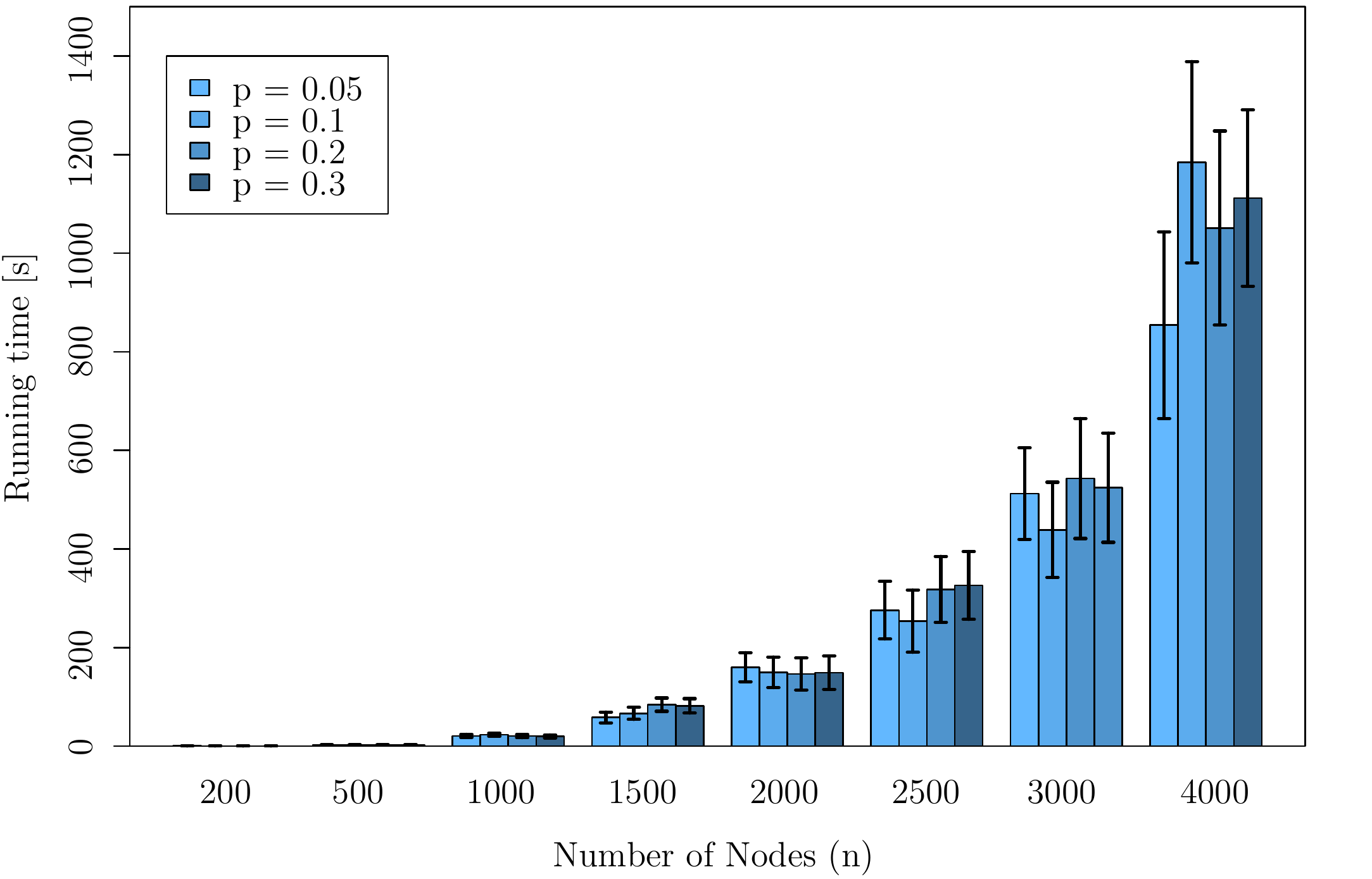}
	\caption{Running time of ILP~\ref{ilp:inter-rack} for different problem sizes. Errorbars indicate the confidence intervals for a confidence level of 0.95.}
	\label{fig:ILPtimes}
	%figure mittels java nonZeroSolver/solver.java und R nonZeroSolver/auswertung.R (beide Dateien im Eclipse)
\end{figure}

		We use ILP~\ref{ilp:inter-rack} to compute $G^{ext}$.
		To show that the running time is acceptable for practical instances
		we are conducting the following experiment on an Intel i7~960 CPU running at 3.2~GHz with 24 GB of DDR3 memory.
		The ILP~\ref{ilp:inter-rack} is solved using Gurobi 5.6.
		We generated problem sequences $D^{\mathrm{ext}}$ for different number of nodes $n$ and a fixed rack size $m = 20$. 
		The single $d^{\mathrm{ext}}_i$'s are
		drawn uniformly at random from the set $\{0, 1, ..., \lfloor n \cdot p \rfloor\}$ for values of $p \in \{0.05, 0.1, 0.2, 0.3\}$.
		For each parameter pair ($n$, $p$) we solved 40 problem instances and measured the running times. 
		Figure~\ref{fig:ILPtimes} plots the required running time against the
		size of the problem instance $n$ and the different values for $p$. One can see that the running time is exponential in $n$ and that
		the number of edges (controlled by parameter $p$) has no significant influence on the running time.
%		Unfortunately, computing exact solutions for data centers with high numbers of nodes using ILP~\ref{ilp:inter-rack} takes too long to be practical.
%		In this cases the polynomial time algorithms relying on finding perfect matchings \cite{Erdoes:arXiv1301.7523, milena02}
%		will be faster (but still far too slow to be practical).

	\subsection{Generating a Traffic Matrix}
	\label{sec:ILPs}
		The process of creating a traffic matrix can be divided into the two steps 
		a) finding the positions of non-zero traffic matrix entries, and
		b) assigning traffic volumes to non-zero traffic matrix entries.\\
		We are finding the positions of non-zero traffic matrix entries by creating a graph $G$ with given intra- and inter-rack node degrees.
		To this end, two graphs $G^{int}$ and $G^{ext}$ are constructed. 
		$G^{int} = (V, E^{int})$ only contains intra-rack edges with degree sequence $D^{\mathrm{int}}$ and
		$G^{ext} = (V, E^{ext})$ only contains inter-rack edges with degree sequence $D^{\mathrm{ext}}$. 
		We construct $G$ by setting $G = \left(V, E^{int} \cup E^{ext}\right)$.
		Whenever there is an edge between a pair of nodes $i$ and $j$ we make $(i,j)$ a random variable in the traffic matrix which is distributed 
		according to  \bytes{PL}{inter} resp. \bytes{PL}{intra}.

		To create $G^{int}$, each rack can be examined separately using Algorithm~\ref{algo:1}.
		This leads to $k$ unconnected subgraphs which model the communication between servers in the same racks.
		However, this only works if the degree sequences are realizable.
		But, as we draw the degree sequences randomly from the given distribution and the rack sizes are relatively small,
		in most cases the demanded degree sequences are not realizable.
		This is problematic as we are not allowed to redraw the degree sequences in that case because this would
		lead to a wrong distribution of intra-rack node degrees.
		Note that this problem is specific to the intra-rack case where the number of nodes is very small and the demanded node degrees are
		very high. 
		For the inter-rack case,
		the probability of sampling a non-realizable degree sequence is much lower because of the large number of nodes and the relatively
		small demanded node degrees. 
		
		To compute intra-rack edges with degrees following \partners{obs}{intra},
		we developed ILP~\ref{ilp:intra-rack}.
		ILP~\ref{ilp:intra-rack} assigns penalties to each node in case it does not meet its demanded node degree.
		The penalty of a node is defined as the absolute difference between the demanded node degree (given by $D^{\mathrm{int}}$) 
		and the node degree in the solution calculated by the ILP itself.
		ILP~\ref{ilp:intra-rack} minimizes the sum over the penalties of all nodes $i$
		divided by the probability of degree $d_i$ (according to \partners{obs}{intra}).
		This way, the sum of the \emph{relative distances} between the degree distribution computed by the ILP and
		\partners{obs}{intra} is minimized.
		
		\begin{ilp}{Constructing an Intra-Rack Graph}
			\begin{flalign*}
				& \text{minimize} & \sum_{0 <i < n} \frac{p_{i}}{Pr(d_i)}  & & \\
				\text{w.r.t.} & ~ & ~ & & \\
				& p_i \geq & 0~~~~~~~~~~~~~~~~~~						&	& \forall 1 \leq i \leq n \\	
				& p_i \geq & \sum_{j \in \intra(i)} b_{i,j} - d_i 	&   & \forall 1 \leq i \leq n \\	
				& p_i \geq & d_i - \sum_{j \in \intra(i)} b_{i,j}   	&   & \forall 1 \leq i \leq n
			\end{flalign*}
			\label{ilp:intra-rack}
		\end{ilp}
		
		In ILP~\ref{ilp:intra-rack}, $p_i$ is the penalty assigned to node $i$. The demanded node degree of $i$ is denoted $d_i$ and
		$b_{i,j}$ is 1 if there is an edge between nodes $i$ and $j$ in the calculated graph.
		For practical instances, the run time of the ILP~\ref{ilp:intra-rack} is not critical as each rack can be examined separately 
		and racks typically consist of up to 40 servers only.
		
\begin{figure}
	\centering
	\includegraphics[scale=1]{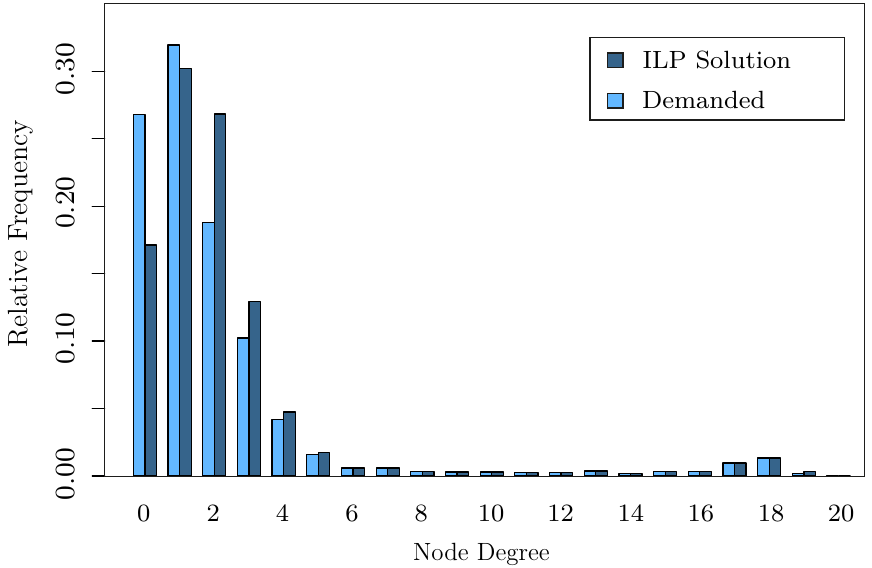}
	\caption{Comparison of a) the distribution of node degrees in the demanded degree sequence and b) the distribution of node degrees of the
	solution computed by ILP~\ref{ilp:intra-rack}}

	\label{fig:ILPcomp}
	%figure mittels ./makePlot.r erstellt (liegt auch in dem verzeichnis)
\end{figure}

		Figure~\ref{fig:ILPcomp} shows the solution quality of ILP~\ref{ilp:intra-rack} in an experiment with 10.000 servers organized in racks 
		of 20 servers each. 
		The 10.000 demanded node degrees are distributed according to \partners{obs}{intra}.
		One can see that the distribution of node degrees computed by ILP~\ref{ilp:intra-rack} is comparable to \partners{obs}{intra}.
		The distributions match very well for node degrees with small densities.
		For larger densities, the gap between the distribution of demanded degrees and the distribution of
		node degrees computed by ILP~\ref{ilp:intra-rack} is larger.
		However, the solution of ILP~\ref{ilp:intra-rack} is of sufficient quality for our purpose.

%	\subsection{Randomizing the Communication Graph}
%		The computed graph $G$ can have artificial patterns 
%		in the structure of the edges caused by the algorithm used to solve our proposed ILPs.
%		To create a graph $G^\prime$ with the same degree sequence as $G = (V,E)$ but without any artificial patterns, we randomize the edges of $G$.
%		To this end, we iteratively pick two edges $(u,v)$ and $(k,l)$, remove them from $E$ and create the two new edges
%		$(u,l)$ and $(v,k)$. We call such an operation an \emph{edge swap}.
%		Note that an edge swap does not distort the degree sequence of the graph.
%		
%		There are two requirements when selecting edges.
%		An edge swap of $(u,v)$ and $(k,l)$ creates the new edges $(u,l)$ and $(k,v)$. Thus, these two edges must not already exist beforehand.
%		The second requirement when swapping inter-rack edges is that neither $u$ and $l$ nor $k$ and $v$ must reside in
%		the same rack. Otherwise the edge swap operation would swap inter-rack edges with intra-rack edges distorting the probability distributions.
%		
%		
%		We randomize each intra-rack graph and inter-rack graph separately and conduct $\alpha s$ \fxnote{Das ist Mist. Ich mache das $n^2$ mal und hoffe für das beste... aber das will ich nicht schreiben.}
%		edge swaps in each graph where $s$ is the number of ,,swappable'' edges.
%		Thus, the probability of not replacing a specific edge is
%		$(1 - \frac{2}{s})^{\alpha s}$. For $\alpha \geq 4$ and $s \geq 100$ this probability is negligibly small.
%		

\section{Generating TCP Flows}
\label{sec:flows}
\subsection{Overview}
	A traffic matrix computed by the Traffic Matrix Creator states the amount of bytes exchanged by node pairs in a fixed time.
	Data-center traffic consists mostly of short-lived flows \cite{MSR-datacenters, datacentersInTheWild}.
	Thus, for each communicating node pair (non-zero TM entry), the bytes have to be separated into different flows.
%	A flow is a series of packets between a source and a destination that logically belong together. 
	We describe a \lf{} flow as the 4-tuple \emph{(start~time, source, destination, size)}.
	This section only deals with payload flows.
	
	Given a \pltm{} determining how many bytes to transfer between every pair of nodes,
	the question answered in this section is: 
	How to separate the non-zero entries of a \pltm{} into flows such that flow sizes are following \size{PL} and 
	the flow inter-arrival times are distributed according to \iat{PL}?
	
	Our strategy is to first generate a set of flows complying with \size{PL} and \iat{PL} and afterwards map these flows
	to the non-zero entries of the \pltm{}.

\subsection{Generating Flows}

Generating flows complying with \size{PL} and \iat{PL} for a given traffic matrix is a challenging task.
A simple approach would be to go through all non-zero TM pairs $(u,v)$ and generate flows for them according to \size{PL} and \iat{PL}.
But this approach raises some questions, for example:

\begin{description}
\item[\emph{When to stop generating new flows for $(u,v)$?}]~\\
We could stop assigning flows to $(u,v)$ when the sum of flow sizes for $(u,v)$ is larger than specified by the TM. But then,
more traffic would be generated than is specified by the TM.
Another way would be to stop generating flows for $(u,v)$ when the next flow that is to be generated would exceed the amount stated by the TM.
This way, the traffic generated by the flows would be less than specified by the TM.
Hence, no matter how we decide, the resulting \gentm would not follow \bytes{PL}{inter} and \bytes{PL}{intra}.
\item[\emph{What if for a small TM entry a huge flow size is generated?}] ~\\
Generating a new flow size in this situation distorts 
the resulting flow-size distribution.
And by assigning the too large flow, the resulting \gentm would not comply with \bytes{PL}{inter} and \bytes{PL}{intra}.
\end{description}
So generating flows for each host pair individually is not practical.

One way to get around these issues is to first create the TM and then a set of ``unmapped'' flows following \size{PL} and \iat{PL} 
(where ``unmapped'' means the flow is not yet assigned to a source-destination pair, \emph{s-d pair}).
Afterwards, flows get mapped to s-d pairs such that the sum of flow sizes mapped to each s-d pair matches
the amount given by the traffic matrix.
However, this mapping has to be done very carefully. 
Since there is no information known about inter-flow dependencies,
the mapping must not introduce any artificial patterns to the generated traffic 
(such a pattern could, for example, be a higher probability to map large flows to node pairs with large TM entries).
Thus, the goal is a random assignment of flows to host pairs $(u,v)$ where the amount of traffic given by the flows between $u$ and $v$ is equal 
to the TM entry $(u,v)$.
We call such a mapping an \emph{exact mapping}.
Note that it is not guaranteed that an exact mapping exists.
Nevertheless, a good mapping strategy assigns flows such that the sum of flow sizes between nodes $i$ and $j$ is as close as possible to TM entry $(i,j)$.

To create flows, we first determine the overall required traffic $s_{M}$ of the TM (as the sum of all entries) 
and then create a set of unmapped flows such that flow sizes sum up to $s_{M}$. We denote the sum of all generated flow sizes as $s_{F}$.
As $s_M$ is a random variable it will hold that $s_{M} = s_{F} \cdot \varepsilon$, $\varepsilon \in \mathcal{R}^+_0$,
where $\varepsilon$ is the imbalance factor between the size of the flows and the TM.
Of course, $\varepsilon$ should be very close to $1$ (meaning there is no imbalance at all), which is why we start over to generate
the \emph{whole set} of unmapped flows with adjusted flow inter-arrival times as long as
$\mid \varepsilon - 1 \mid > 0.01$. This means that the sum of all generated flow sizes deviates at most 1\% from the traffic specified by the TM.
We assume this to be a reasonably small error.
%To account for the variations from the random processes, we scale the arrival times of the flows to match the sum of the flow sizes 
%to the sum of traffic from the TM.
%Otherwise it could happen that the traffic defined by the unmapped flows does not match the overall required traffic specified by the TM.

We will now present two different strategies to map the unmapped flows to node pairs. The first one is a purely random process and the second one uses a variation of the queuing strategy
\emph{deficit round robin} (DRR) \cite{shreedhar1995efficient}. Afterwards, we study the quality of both strategies.

The randomized assignment uses the TM as a probability distribution and, for each generated flow, draws a node pair from this distribution.
In this process, we define the initial probability to assign a flow to node pair $(i,j)$ as the TM entry $(i,j)$ divided by $s_{M}$.
After a flow has been assigned, the probability distribution at the point of the node pair is lowered proportionally to the size of the flow.

The second strategy is inspired by DRR. DRR schedules jobs of different sizes and classes onto a shared processor.
The goal of DRR is to share the processor among all classes according to the ratio of their priorities. 
To this end, each class is assigned a \emph{priority} and a \emph{credit}. 
DRR loops Round Robin through all classes. 
In each iteration of the loop, the credit of each class is raised by some constant (called quantum) weighted by the priority of the class.
If for a class there exists a job with a size smaller than the current credit of the class,
this job is scheduled to the processor and the credit of the class is lowered by the size of the job.

We use a DRR variant to map flows to node pairs. 
In this variant, node pairs correspond to classes and
flows correspond to jobs. The only difference in our variant is that we do not schedule flows onto a shared processor; we schedule flows on node pairs.
The priority of a node pair is proportional to the size of its residual traffic matrix entry.
We loop Round Robin over all node pairs and raise their credit proportional to their residual TM entry.
Whenever the unmapped flow under consideration is smaller then or equal to the credit of the node pair, this flow is mapped to the node pair and the 
credit is lowered accordingly.

\begin{figure}[t]
			\centering
			\includegraphics[scale=0.85]{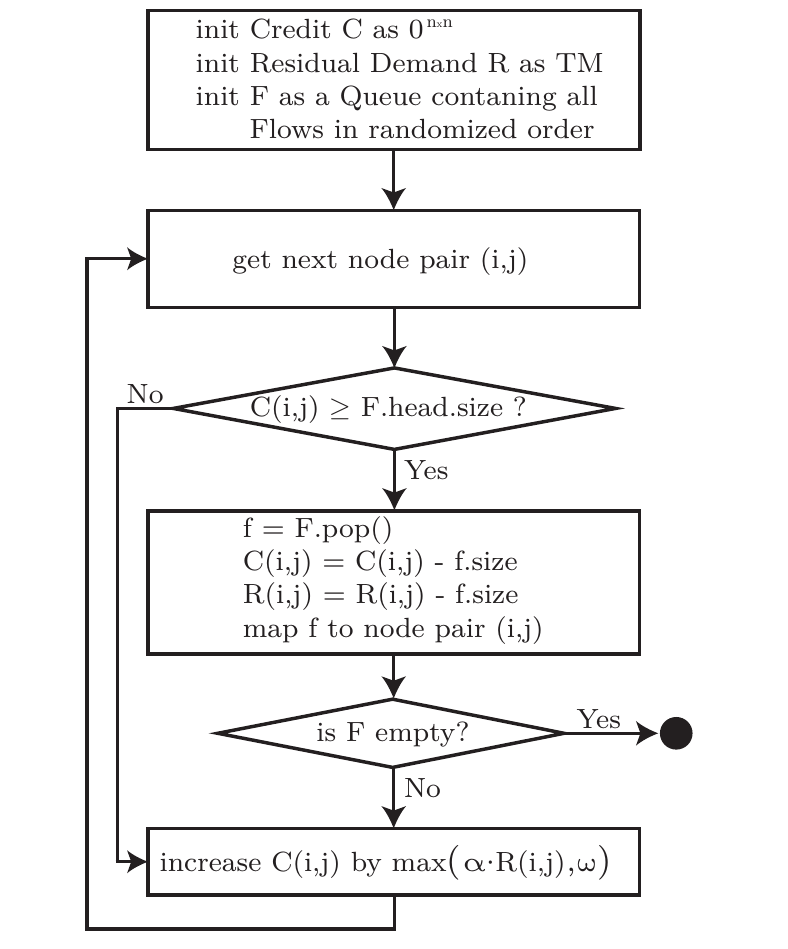}
	        \caption{Deficit Round Robin inspired algorithm for selecting s-d pairs for flows.}
    		    \label{alg:srcdst}
		\end{figure}

Our adapted version of the DRR strategy can be seen in Figure~\ref{alg:srcdst}. 
In this algorithm, $i$ always corresponds to a source, $j$ to a destination and
$R$ is the \emph{residual} traffic between $i$ and $j$ as specified by the TM:
whenever a flow is assigned to $(i,j)$, $R_{i,j}$ is decreased by the size of the flow.
$F$ is a queue that initially contains all flows in a randomized order.
$C_{i,j}$ is the credit (akin to DRR) of the node pair $(i,j)$. 
$C_{i,j}$ is decreased whenever a flow is assigned to $(i,j)$
by the size of the flow. 
The algorithm iterates Round Robin over all node pairs and tries to assign the flows queued in $F$.
For each flow $f$ the algorithm iterates as long over the node pairs $(i,j)$ as no valid candidate has been found. 
$(i,j)$ is a valid candidate for flow $f$ if $C_{i,j}$ is larger than or equal to the size of $f$.
After a pair $(i,j)$ has been inspected its deficit counter is increased by $\max(\alpha \cdot R_{i,j}, \omega)$;
$\alpha$ and $\omega$ control the increase of the deficit counter over time.
Ideally, both parameters are chosen to be very small. 
We found that setting them to values below $\alpha=0.1$ and $\omega=100$ cause no significant improvement of the flow assignment and 
only increases the run time of the algorithm. Thus, we consider $\alpha=0.1$ and $\omega=100$ to be a good choice.\footnote{Faster implementations without looping are obviously possible and not difficult. 
We concentrate on the core idea here since this is not a time-critical aspect.}

\subsection{Quality of Flow Assignment}

In an optimal flow assignment, each node pair is assigned flows which
exactly sum up to the amount of traffic stated by the given TM.
In reality, we will produce a traffic matrix with slight derivations.
To express the difference between the given TM $M$ and the TM $M^\prime$ produced by the flow assignment
we interpret both $M$ and $M^\prime$ as probability distributions of exchanging traffic.
Then, we express the distance between these two distributions by the relative entropy.
% we use the 
%measure \emph{relative entropy} which expresses the difference between two probability distributions.
%This makes sense because a TM can also be interpreted as a probability of exchanging traffic. 
%So we use $M$ as the ground truth and express the difference between $M$ and $M^\prime$ as the entropy of both matrices.
The relative entropy is naturally defined as the \emph{Kullback-Leibler divergence} (KL), but KL requires that 
$M^\prime_{i,j}=0~\Rightarrow~M_{i,j}=0~\forall~(i,j)\in n \times n$, which does not hold in our case. 
However, the symmetric form of KL, called \emph{Topsøe distance} (Equation~\ref{equ:topsoe}) \cite{johnson2001symmetrizing}
does not require this implication and can be used instead to compute the distance between two probability distributions.

\vspace{0.3cm}

Topsøe$(M, M^{\prime})  =$
\begin{eqnarray}
	\label{equ:topsoe}
	\sum_{(i,j)} 
		\left(	
			M_{i,j} \ln \frac{2 M_{i,j}}{M_{i,j} + M^{\prime}_{i,j}} + 
			M^{\prime}_{i,j} \ln \frac{2 M^{\prime}_{i,j}}{M_{i,j} + M^{\prime}_{i,j}} 
		\right)
\end{eqnarray}

We look at the Topsøe distance for \emph{different} load levels of a network because
given a fixed flow-size distribution, an increasing communication volume (TM size) will influence the results of the flow assignment methods:
If the total traffic volume tends towards infinity, a single flow gets very small compared to a TM entry.
In such a scenario it is very easy to find matching flow assignments.
A load level is created by multiplying the TMs with a factor $l$;
we denote the corresponding TM with $lM$.
We then assign flows for $lM$ to s-d pairs and calculate the TM $(lM)^{\prime}$ based on that flow assignment.
%The entry $(i,j)$ of $M^{(l)\prime}$ is the sum of the sizes of all flows from $i$ to $j$.

\begin{figure}
			\centering
				\includegraphics[width=0.49\textwidth]{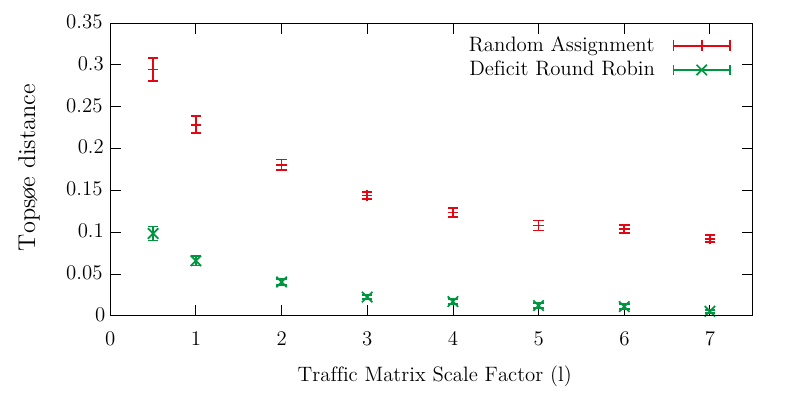}
				\caption{Topsøe distance of flow assignment methods over different traffic volumes. Error bars show confidence intervals for a confidence level of 95\%.}
				\label{fig:entropy}
		\end{figure}

We use $lM$ as the ground truth and express the difference between $lM$ and $(lM)^{\prime}$ as the relative entropy of both matrices.
Figure~\ref{fig:entropy} shows the relative entropy obtained via either the random strategy or the Deficit Round Robin strategy
calculated as averaged over the \emph{Topsøe distance} of 40 matrices of 10\,s generated traffic each
$\left( \frac{1}{40} \sum_{i=1}^{40} \text{Topsøe}\left( lM, \left(lM\right)^{\prime} \right) \right)$.
The data center for which we generate traffic consists of 75 racks with 20 servers each.
It is the same size that was used in the study \cite{MSR-datacenters}.
It can be seen that for both methods the Topsøe distance decreases with increasing load but for Deficit Round Robin the relative entropy
is much lower, thus the method achieves a better flow assignment than the random mapping process.
We will only consider the DRR-based scheme henceforth.

\section{Empirical Evaluation}

\label{sec:evaluation}
	\subsection{Approach}		
%		For the evaluation of our traffic generator we are using both a statistical analysis of our generated TCP schedule and
%		a network emulation using MaxiNet~\cite{wette14b}.
%		The goal of our traffic generator is to compute a schedule of payload transmissions which (when transported using TCP) produces traffic
%		following the six probability distributions given beforehand.
%		To show that our generated traffic reproduces the,
%		we are using a twofold analysis on the generated traffic to determine all necessary statistical properties of the resulting traffic.

		\genname{} works properly if it is able to compute a schedule of TCP payload transmissions where (when transferred over a network)
		a) the generated \gentm{} has the same properties as \obstm{} and 
		b) the generated flows have the same properties as the observed flows.
		
		We are using a stochastic analysis of the generated TCP schedule to confirm that \gentm{}
		follows the same probability distributions as \obstm{}.
		To this end, we compute 
		\partners{gen}{intra}, \partners{gen}{inter}, \bytes{gen}{intra}, \bytes{gen}{inter}, and \iat{gen}
		based on the \lf{} schedule and compare them to
		\partners{obs}{intra}, \partners{obs}{inter}, \bytes{obs}{intra}, \bytes{obs}{inter}, and \iat{obs}.
		A network emulation is used to capture the effects of TCP when the generated traffic is replayed on a data-center topology.
		From the results of the emulation, we compute \size{gen} and compare them to \size{obs}.
		The emulation environment we are using is MaxiNet \cite{wette14b}.

	\subsection{Traffic Properties used in the Evaluation}
According to \cite{MSR-datacenters}, \partners{obs}{intra} and \partners{obs}{inter} are heavy-tailed in typical data centers. 
It is reported that for a pair of
servers located in the same rack, the probability of communicating in a fixed 10\,s period is 11\,\% whereas the probability for out-of-rack 
communication for any pair of servers is only 0.5\,\%. 
In addition, a server either talks to the majority of servers in its own rack or to less than one forth of them.
The amount of traffic that is exchanged between server pairs is distributed based on their relationship: 
Servers in the same rack either exchange only a small amount or a large amount of data, 
whereas traffic across racks is either small or medium per server pair. 

Kandula et al. \cite{MSR-datacenters} found that 80\,\% of the flows in the data center last no longer than 10\,s and that only 0.1\,\% of the flows last longer than 200\,s.
More than half the traffic is in flows shorter than 25\,s and every millisecond 100 new flows arrive at the network.

An independent study \cite{datacentersInTheWild} looked at traffic from 10 different data centers. They
showed that across all 10 data centers \size{obs} is nearly the same. Most of the flows were smaller than 10\,KB and
10\,\% of the flows are responsible for more than half of the traffic in the data centers.

For evaluation, we used the observed distributions by \cite{MSR-datacenters, datacentersInTheWild} as an input to our traffic generator.
Both studies reason about all the traffic in data centers. In addition to traffic transported with TCP, this includes ARP, DNS and many more
protocols that do \emph{not} use TCP for transport.
This results in traffic characteristics that cannot be reproduced using TCP only. 
A flow resulting from an ARP request, for example, has a size of 66\,Bytes which was also the smallest reported flow size.
Due to the three-way handshake used to establish and tear down TCP sessions the smallest possible flow size (on Layer~2) TCP can produce is 272\,Bytes.
For evaluation we increased all flow sizes by 219\,Bytes to remove this mismatch.

As a result of that increase, 
\bytes{obs}{intra} and \bytes{obs}{inter} no longer match the enlarged \size{obs}.
This makes it impossible to have a good flow assignment because there are not enough small flows to be mapped to the small non-zero TM entries.
To counteract this, we increased \bytes{obs}{intra} and \bytes{obs}{inter} by 1000\,Bytes.

Note that the performed changes are only minor.
The average flow size extracted from \cite{datacentersInTheWild} is 142\,KB.
Thus, increasing the size of each flow by 219\,Bytes is an increase of 0.15\,\% on average.
The average non-zero intra-rack traffic matrix entry has a size of 12.6\,MB, the average non-zero inter-rack traffic matrix entry 12.4\,MB.
Thus, an increase of 1000\,Bytes per non-zero traffic matrix entry is negligible (about 0.1 \%).

\subsection{Topology and Emulation Environment}
%We use MaxiNet~\cite{wette14b} to emulate a mid-sized data center network.
%MaxiNet is an open source toolkit which distributes Mininet \cite{mininet} emulations to multiple machines.
%This gives us the opportunity to emulate large data center networks although these networks would be too large to be emulated on a single Mininet
%instance.

\begin{figure}
			\centering
				\includegraphics[width=0.49\textwidth]{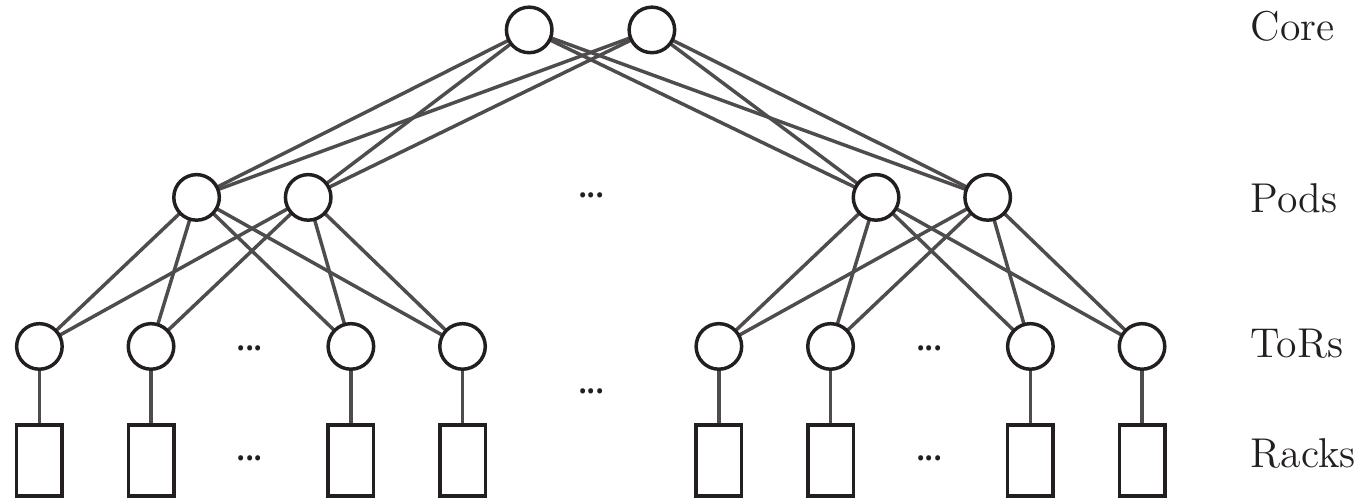}
				\caption{Sketch of the Clos-like topology that was used in our experiments.}
				\label{fig:clos}
		\end{figure}

To include the effects of TCP into our evaluation,
we choose to emulate a data center consisting of 72 racks employing a \emph{Clos-like topology}.
From the emulation, we are able to determine \size{gen} and \iat{gen}.
A sketch of the emulated topology can be seen in Figure~\ref{fig:clos}. 
Each rack consists of 20 servers and one ToR switch, which makes 1440 servers overall.
Servers are connected by 1\,Gbit/s links to ToR switches.
Pods consist of \emph{eight} ToR switches which are connected to \emph{two} pod switches with 10\,Gbit/s links.
Pod switches are connected to two \emph{core switches} with 10\,Gbit/s links.
The core layer in our topology consists of \emph{two} switches. We assume a forwarding delay of 0.05\,ms per switch.
In each experiment, we emulated 60 seconds of traffic. This traffic was generated from the statistics reported in the previous section.
We used a time dilation factor of 200, which means one experiment completed after 200 minutes.

For emulation, we used 12 physical worker nodes equipped with Intel Xeon E5506 CPUs running at 2.16 GHz, 
12 Gbytes of RAM and 1GBit/s network interfaces
connected to a Cisco Catalyst 2960G-24TC-L Switch.
Routing paths are computed using equal cost multipath (ECMP) implemented on the Beacon controller platform \cite{erickson2013beacon}. 
As the controller was placed out-of-band 
and did not use any kind of time dilation, the routing decisions of the single controller were fast enough for the whole data center network.
In addition, the latency between the controller and the emulated switches was not artificially increased. This means that in relation to 
all the other latencies in the emulated network, the controller decisions were almost immediately present at the switches and did not 
add any noticeable delay to the flows.
Please note that for a real data center (without using time dilation) an ECMP implementation based on only one centralized controller
would likely not keep up with the high flow arrival rates; for details see \cite{wette14b}.

	\subsection{Results}
		To verify that \genname{} produces a reasonable flow schedule, the traffic created by the schedule (box 5 in Figure~\ref{fig:workflow}) 
		must have the same properties as the observed traffic (box 1 in Figure~\ref{fig:workflow}).
		As a) we do not have access to the observed \lt{} traces and b) it is unclear how to directly compare two packet traces with each other,
		we compare the statistical properties of the two traces with each other (boxes 2 and 6 in Figure~\ref{fig:workflow}).
		Comparison is done throughout the following sections where each statistical property is inspected individually.
		Due to the huge amount of samples (our collected packet traces contain 7,060,194 flows, 330,155 distinct intra-rack and 1,675,305 inter-rack TM entries; 
		each of the generated \lf{} schedules contain $\sim$\,6\,M flows)
		it is not easily possible to use any goodness of fit test to judge whether the generated distributions match
		the corresponding observed distributions.
		This is because there exist small statistical differences between both distributions that together with the large set of samples are big enough for the 
		goodness of fit tests to reject, but too small to be of practical importance for our purpose
		(these differences are statistically significant, but not relevant).
		We instead analyze the distributions by using 
		Probability-Probability plots (PP-plots) and Quantile-Quantile plots (QQ-plots)\footnote{In a PP-plot the cumulative probabilities of two distributions are plotted against each other. QQ-Plots are plotting the quantiles of both distributions against each other. If the plot shows the identity function,
		this is an indicator that the distributions are fitting \cite{GraphicalMethodsforDataAnalysis}.}.

		\subsubsection{Generated Flow-Size Distribution}
		\label{sec:GeneratedFlow-SizeDistribution}
			To determine \size{gen}
			we emulated 60 seconds of data-center traffic consisting of 1440 hosts as described previously.
			A packet trace was captured on the first interface of each emulated core switch.
			We conducted 16 independent experiments (with 16 different \lf{} schedules) and used the corresponding 32 traces to compute \size{gen}.
			The number of captured flows over all experiments is 7,060,194. 
			%We binned them by size into bins of 100\,Bytes and compared 
			%the distribution to the targeted distribution of flow sizes.
			
		\begin{figure}
			\centering
				\includegraphics[scale=1]{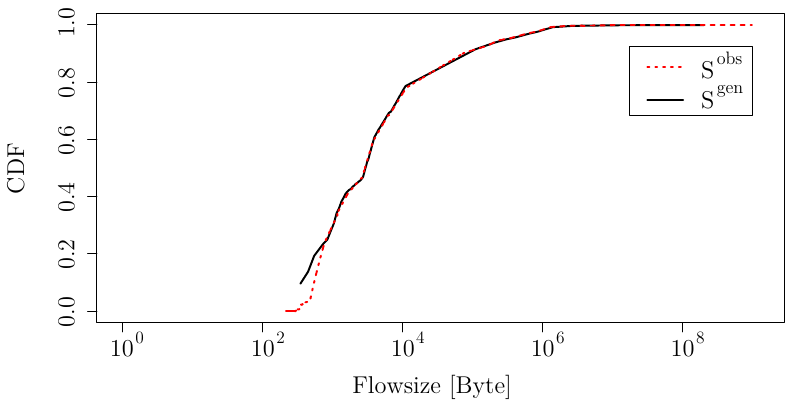}
				\caption{Comparison between \size{gen} and \size{obs}.}
				
				\label{fig:flowSizeDistributionVergleich}
				
			%figure erzeugt mittels: python ~/promotion/forschung/packetSamplingEvaluation/printFlowSizeCDF.py
			%dies erzeugt die datei /tmp/flowSizeCDF.txt
			%diese kann mit folgemdem r-script geplottet werden:
			
			%setwd("~/promotion/forschung/dataCenterTrafficGenerator/")
			%cdf_flowSize <- read.csv("flowSizeShiftedBy219byte.csv", header = FALSE)
			%cdf <- read.csv("/tmp/flowSizeCDF.txt", header=F, sep=",")
			
			%idx = c(seq(2, 100, 1), seq(110, 41000, 1000))
			
			%plot(t(cdf)[idx,1], t(cdf)[idx,2], log="x", ylim=c(0,1), xlim=c(1, 1.25e9), type="l", lwd=2, lty="solid", xlab="Flow size", ylab="CDF")
			%lines(t(cdf_flowSize), col="red", lwd=2, lty="dotted")
			%legend(1.5e6, 0.9, c("Original", "Generated"),text.width=2, lty=c("dotted", "solid"), col=c("red", "black"), lwd=2)

		\end{figure}

			Figure~\ref{fig:flowSizeDistributionVergleich} plots \size{obs} and \size{gen}.
			It can be seen that the distributions clearly match for flow sizes larger 1000\,bytes. 
			The distributions of smaller flows, however, do not match well.
			We suspect this is partly due to the behavior of TCP and partly due to our assumptions on the size of ACK flows
			as most flows smaller than 1000\,bytes are ACK flows (Figure~\ref{fig:convolution}).
			As discussed in Section~\ref{sec:payloadtoackratio}, smaller flows tend to have a lower ACK-to-payload ratio.
			The Flowset Creator, however, calculates the size of each induced ACK flow with a fixed ratio of $r$ which results in 
			the slightly wrong distribution of ACK-flow sizes.

		\subsubsection{Inter-Rack Comm. Partners}
			\begin{figure}
				\centering
					\includegraphics[width=0.49\textwidth]{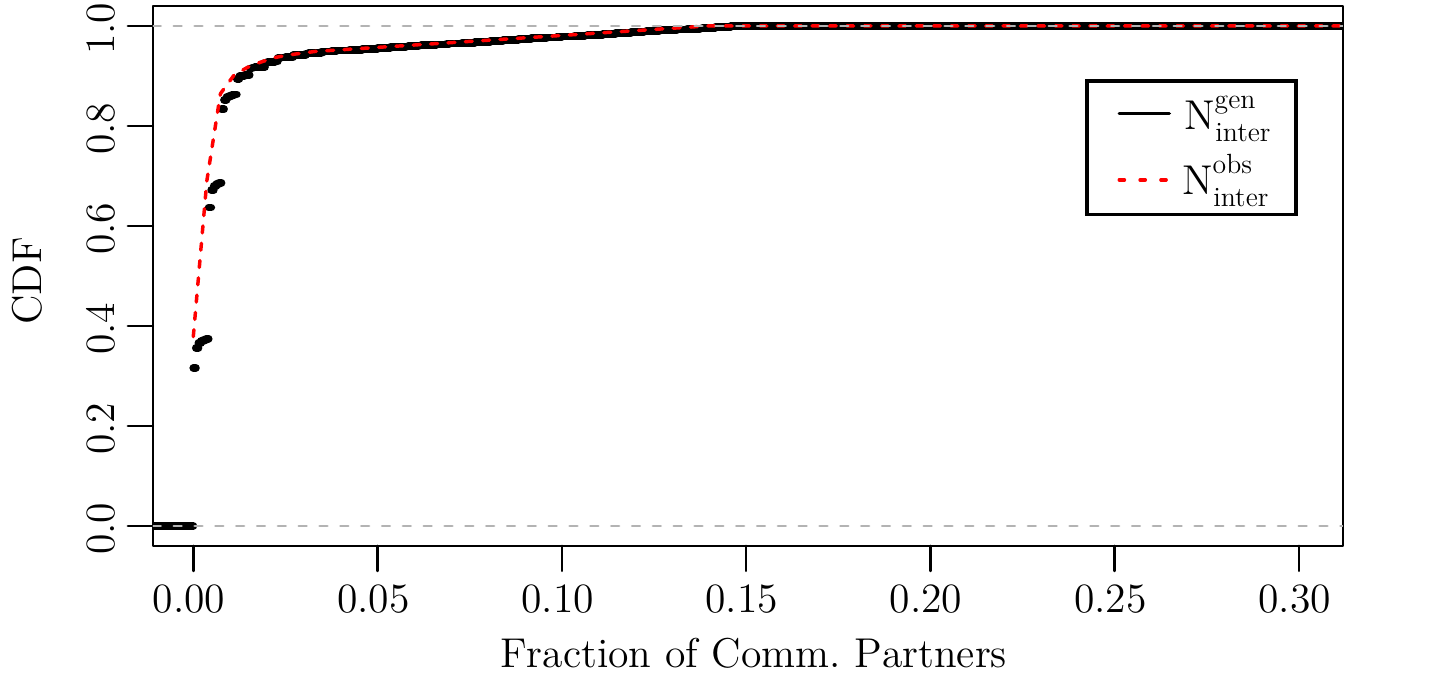}
					\caption{Comparison between \partners{gen}{inter} and \partners{obs}{inter}.}
					\label{fig:orp_plot}
				%erzeugt durch die python scripte in trafficAnalyzer
			\end{figure}
		
			\begin{figure}
				\centering
					\includegraphics[width=0.49\textwidth]{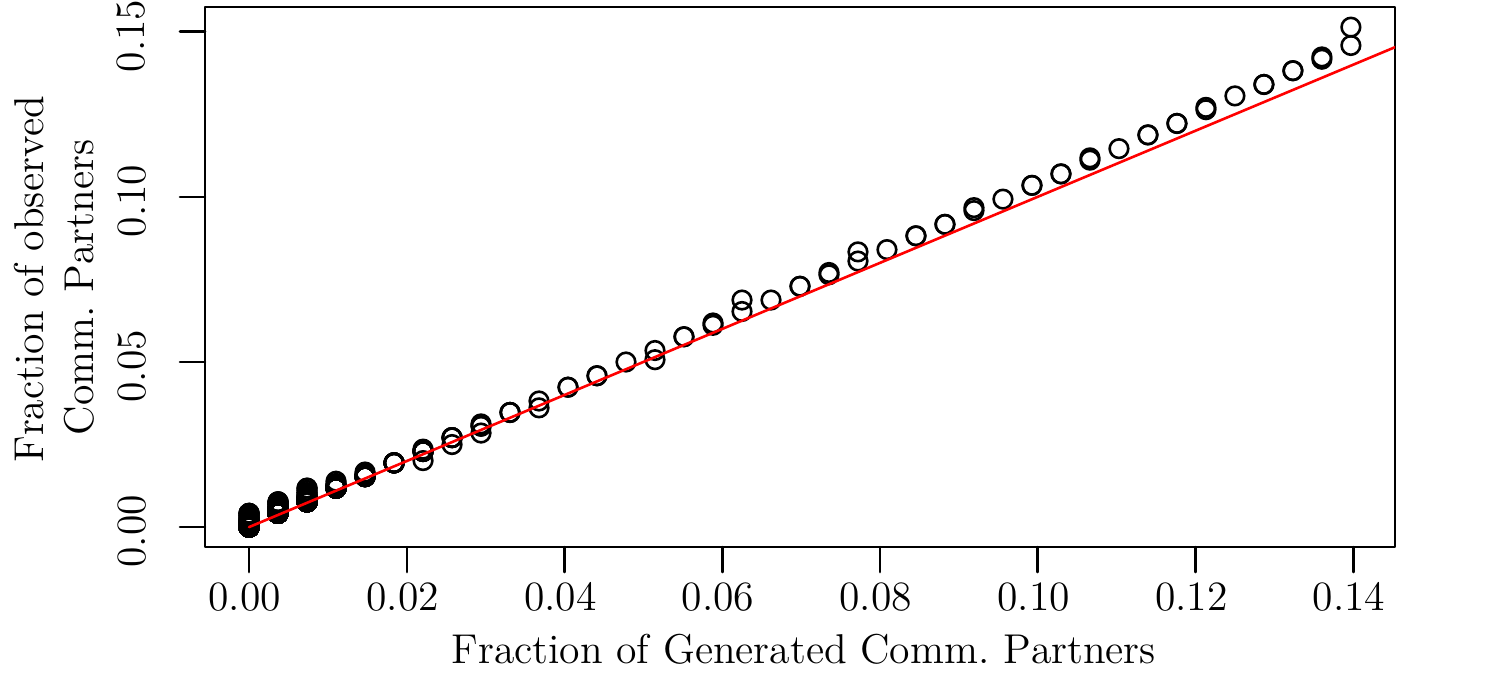}
					\caption{QQ-plot of \partners{gen}{inter} and \partners{obs}{inter}.}
					
					\label{fig:orp_qq-plot}
				%erzeugt durch die python scripte in trafficAnalyzer
			\end{figure}			
			
			To determine
			\partners{gen}{intra} and \partners{gen}{inter}
			we used the same 16 traffic schedules as before.
			%Each schedule contained 60 seconds of traffic for 1440 servers organized in 72 racks of 20 servers each.
			\partners{gen}{inter} and \partners{obs}{inter} are plotted in Figure~\ref{fig:orp_plot}.
			From the plot no difference between the two distributions is discernible.
			The corresponding QQ-plot (Figure~\ref{fig:orp_qq-plot}) also does not show any significant differences
			between \partners{obs}{inter} and \partners{gen}{inter}.

		\subsubsection{Intra-Rack Comm. Partners}
		\label{sec:Intra-RackCommPartners}
		\begin{figure}
			\centering
				\includegraphics[width=0.49\textwidth]{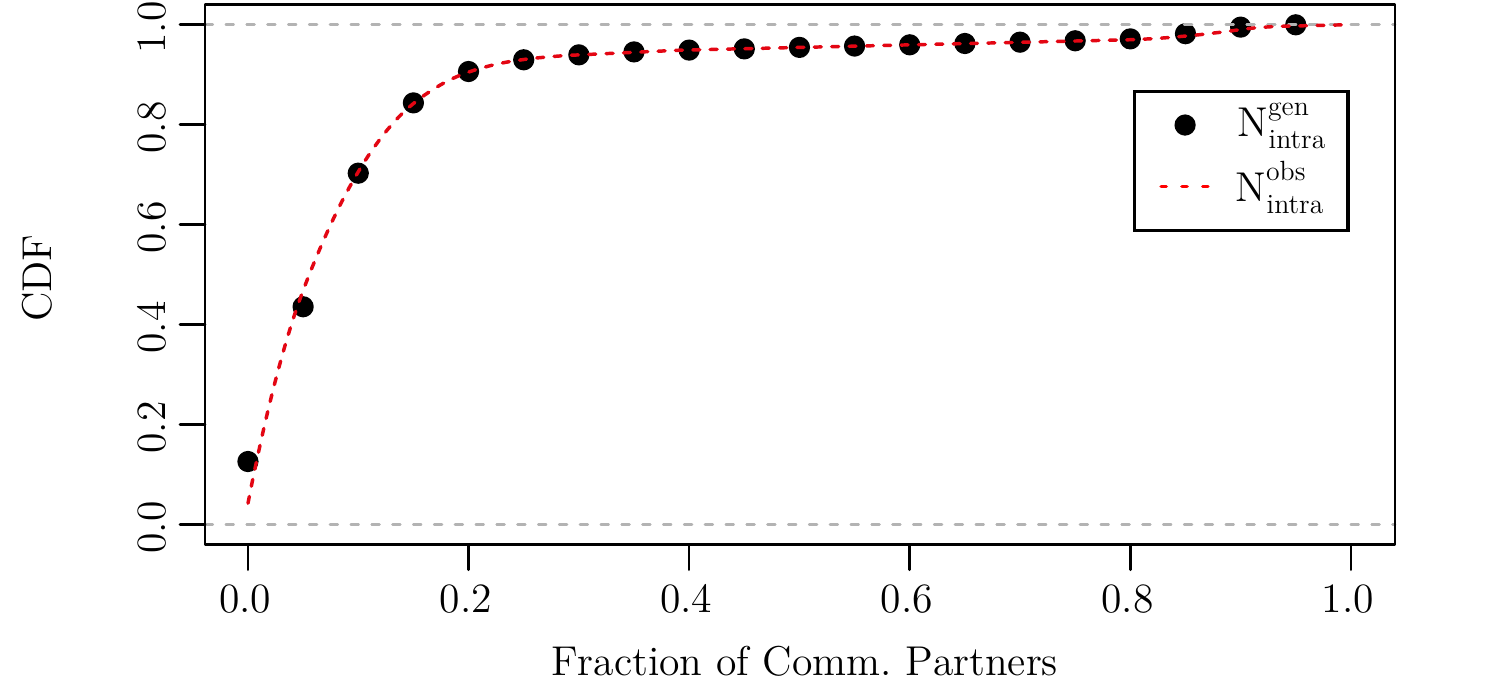}
				\caption{Comparison between \partners{obs}{intra} and \partners{gen}{intra}.}
				\label{fig:irp_plot}
			%erzeugt durch die python scripte in trafficAnalyzer
		\end{figure}
		
		\begin{figure}
			\centering
				\includegraphics[width=0.49\textwidth]{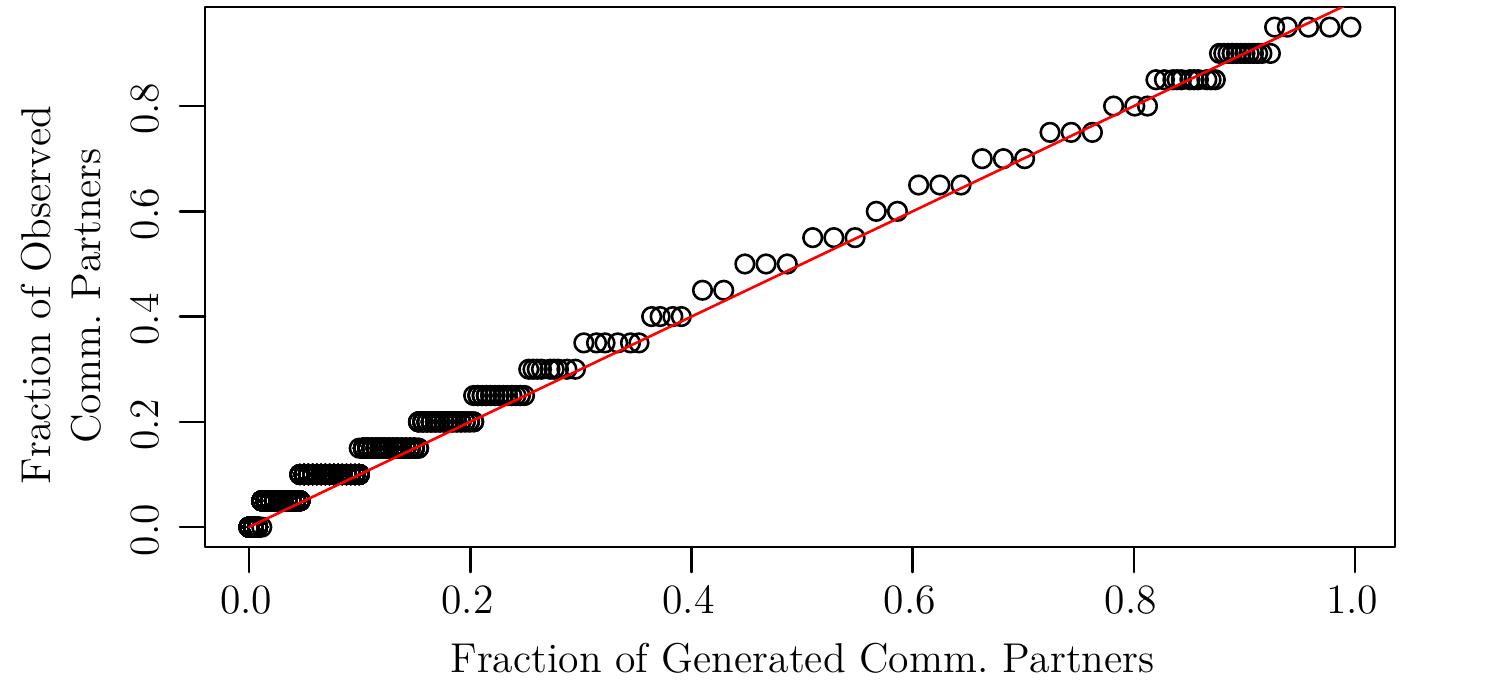}
					\caption{QQ-plot of \partners{obs}{intra} and \partners{gen}{intra}.}
				\label{fig:irp_qq-plot}
			%erzeugt durch die python scripte in trafficAnalyzer
		\end{figure}
		The comparison between \partners{obs}{intra} and \partners{gen}{intra}
		(Figure~\ref{fig:irp_plot})
		shows that our generated traffic contains a little too many intra-rack communication partners with a low degree.
		Despite that, both CDFs are nearly identical. This can also be confirmed by looking at the corresponding QQ-Plot (Figure~\ref{fig:irp_qq-plot}).
		The plot shows an almost straight line that lies a bit above the identity function.
		This result is in line with what is discussed in Section~\ref{sec:ILPs}.

		\subsubsection{Intra-Rack Traffic}
		\begin{figure}
			\centering
				\includegraphics[width=0.49\textwidth]{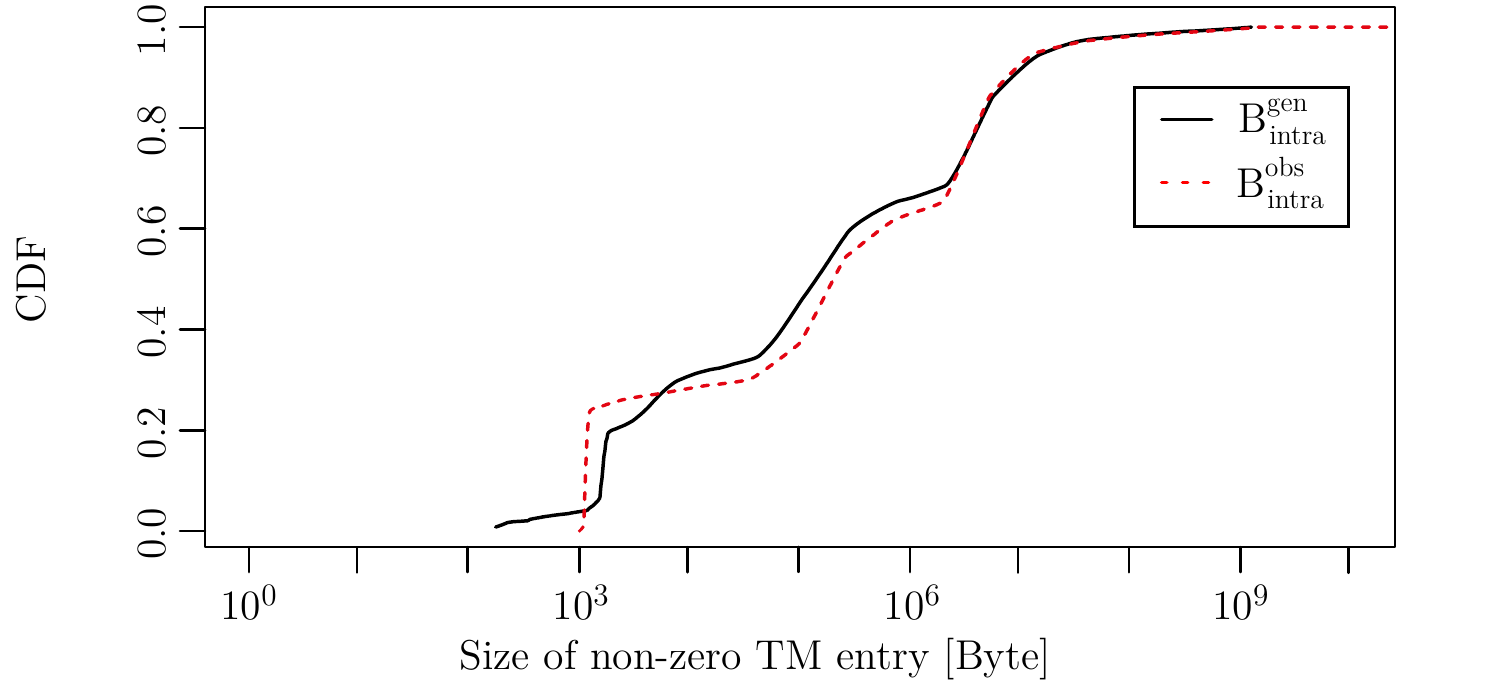}
				\caption{Comparison between \bytes{obs}{intra} and \bytes{gen}{intra}.}
				\label{fig:irt_plot}
			%erzeugt durch die python scripte in trafficAnalyzer
		\end{figure}
		
		\begin{figure}
			\centering
				\includegraphics[width=0.49\textwidth]{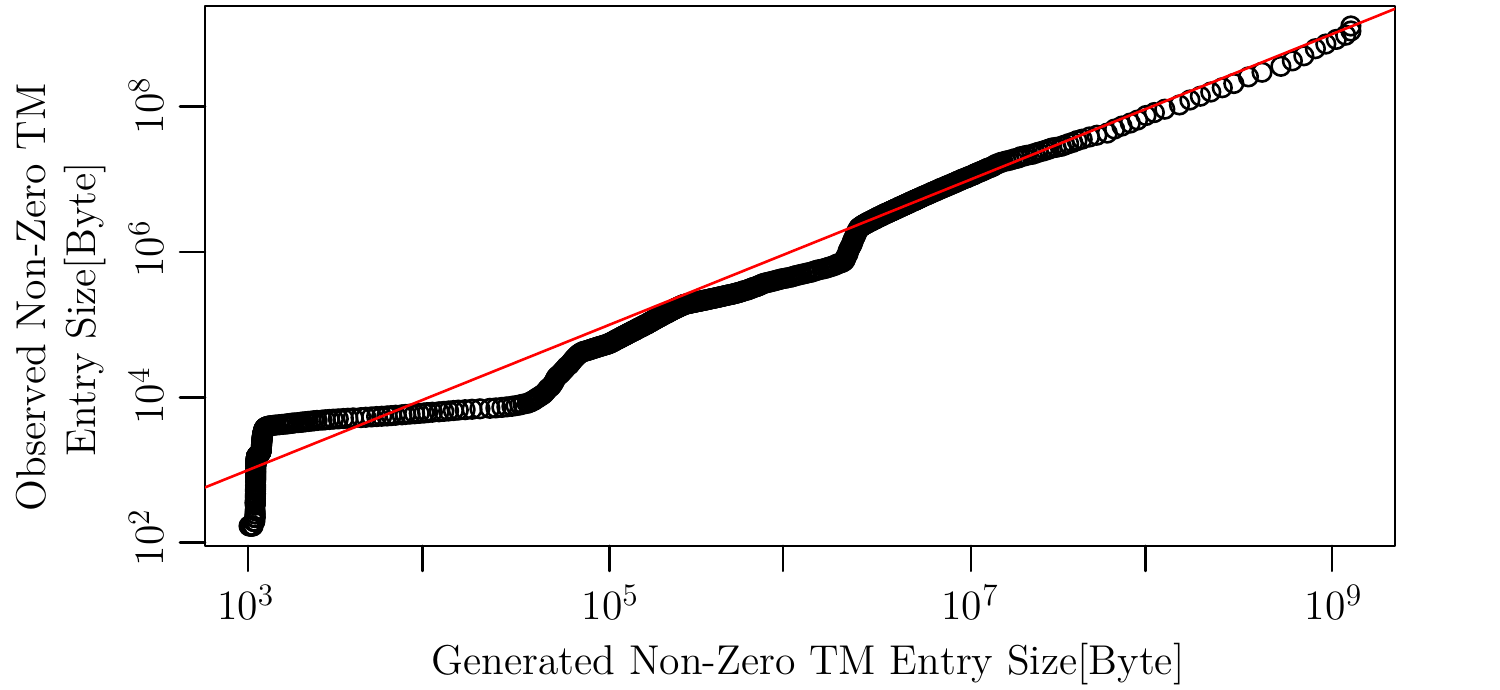}
				\caption{QQ-plot of \bytes{obs}{intra} and \bytes{gen}{intra}.}
				\label{fig:irt_qq-plot}
			%erzeugt durch die python scripte in trafficAnalyzer
		\end{figure}
		The \obstm{}s used in this section are deduced from the same 16 traffic schedules we used
		in Section~\ref{sec:Intra-RackCommPartners}. To compute the single traffic matrix entries, we
		fixed the payload-to-ACK ratio $r$ to $2.5$ (see Section~\ref{sec:payloadtoackratio})
		and computed the size of the flows on Layer~2 between each pair of servers. 
		From that, we calculated the respective 96 \obstm{}s (each for a period of 10\,s).
		
		The corresponding \bytes{gen}{intra}
		is compared to \bytes{obs}{intra} in Figure~\ref{fig:irt_plot}.
		Except for entries smaller than $10^4$ Bytes \bytes{gen}{intra} is strictly following \bytes{obs}{intra}.
		This can further be confirmed by the QQ-Plot (Figure~\ref{fig:irt_qq-plot}) which additionally only shows a small anomaly of the distribution
		for entries around $10^6$ Bytes.
		
		The difference between both distributions in the smaller entries is due to the process of mapping single flows to traffic matrix entries.
		The goal of the Mapper is to distribute flows to traffic matrix entries such that for each node pair the difference between 
		their TM entry and the sum of flow sizes between that nodes is minimized per server pair.
		The smaller the TM entry, the fewer flows can be mapped onto the corresponding node pair which means it is harder to find a well fitting mapping.

		\subsubsection{Inter-Rack Traffic}
		\begin{figure}
			\centering
				\includegraphics[width=0.49\textwidth]{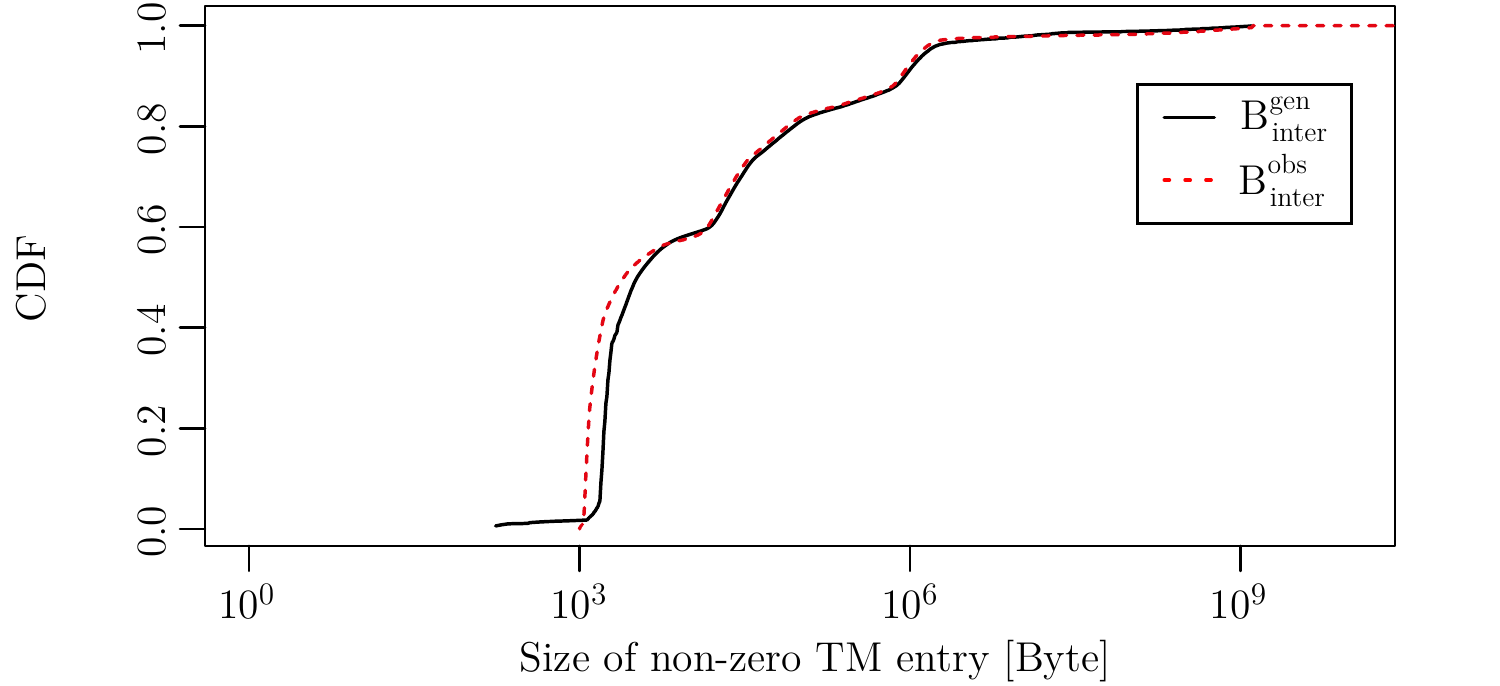}
				\caption{Comparison between \bytes{obs}{inter} and \bytes{gen}{inter}.}
				\label{fig:ort_plot}
			%erzeugt durch die python scripte in trafficAnalyzer
		\end{figure}
		
		\begin{figure}
			\centering
				\includegraphics[width=0.49\textwidth]{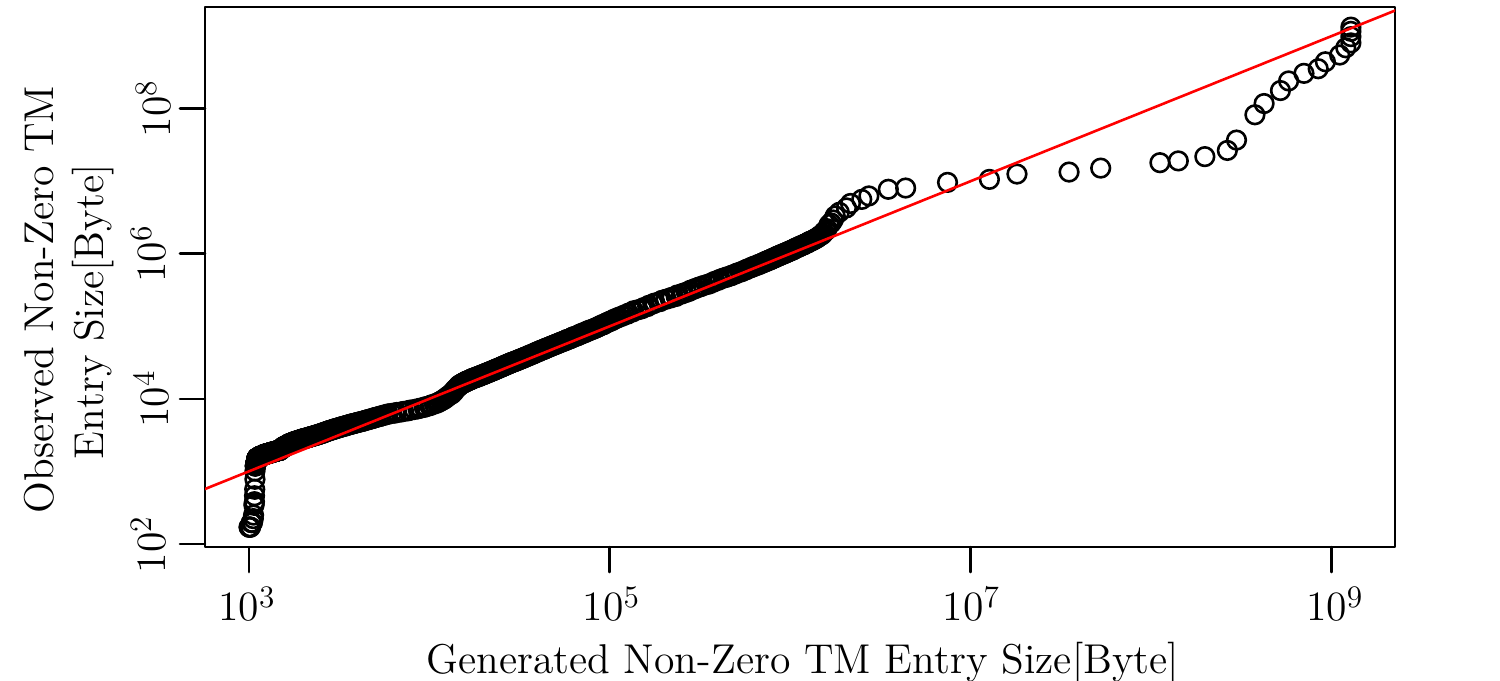}
				
				\caption{QQ-plot of \bytes{obs}{inter} and \bytes{gen}{inter}.}
				\label{fig:ort_qq-plot}
			%erzeugt durch die python scripte in trafficAnalyzer
		\end{figure}
		
		\begin{figure}
			\centering
				\includegraphics[width=0.49\textwidth]{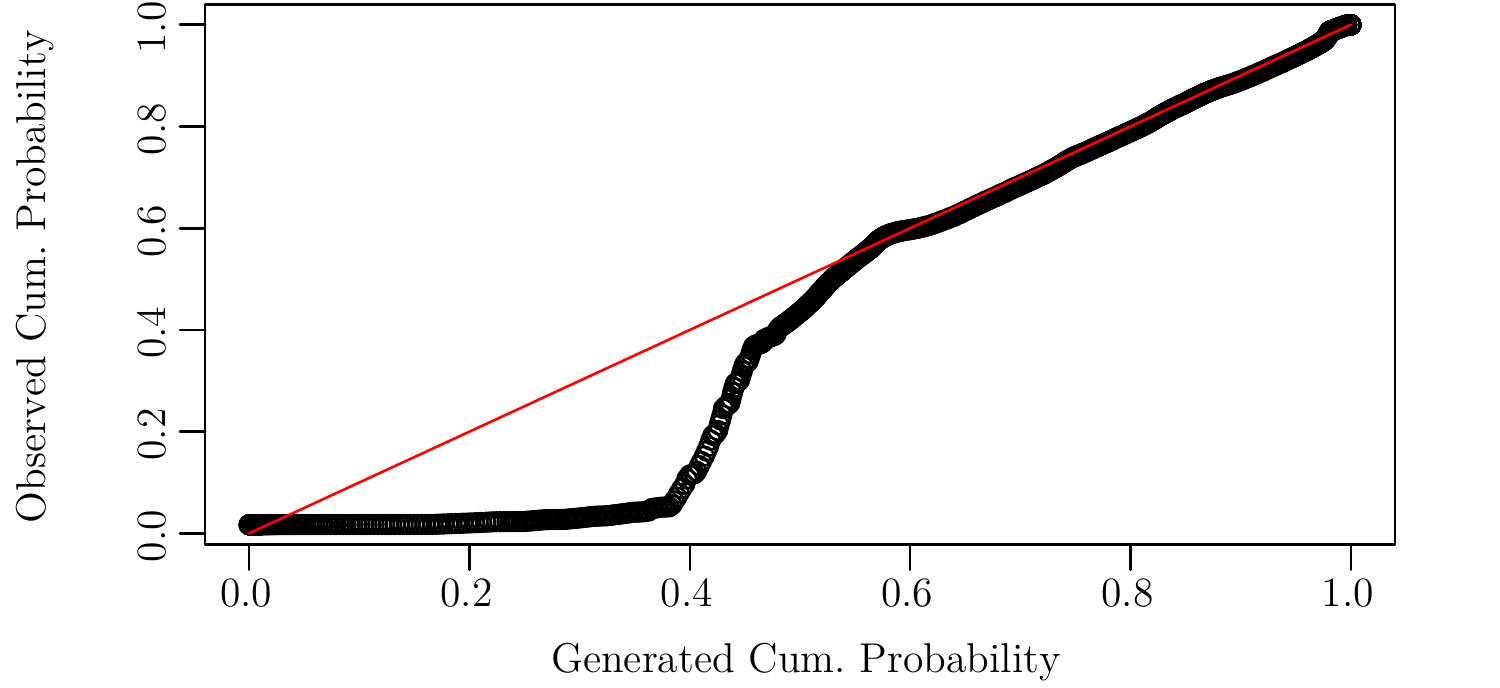}
				\caption{PP-plot of \bytes{obs}{inter} and \bytes{gen}{inter}.}
				\label{fig:ort_pp-plot}
			%erzeugt durch die python scripte in trafficAnalyzer
		\end{figure}
		
		\bytes{obs}{inter} and \bytes{gen}{inter} are plotted in Figure~\ref{fig:ort_plot};
		the corresponding QQ-plot can be seen
		in Figure~\ref{fig:ort_qq-plot}.
		From Figure~\ref{fig:ort_plot}, we observe the same situation as in the
		\emph{intra}-rack case. The QQ-plot additionally exposes differences for the distribution of large entries ($>10^7$).
		This effect in the QQ-plot is caused by only a slight difference between the tails of both distributions. As the tails of both
		\bytes{obs}{inter} and \bytes{gen}{inter} are very long, slight differences in the probabilities have a huge impact on the QQ-plot.
		When consulting the corresponding PP-plot shown in Figure~\ref{fig:ort_pp-plot} we immediately see that 
		the probabilities for the generated traffic and the input to our traffic generator are comparable.
		Thus, \bytes{obs}{inter} and \bytes{gen}{inter} are well fitting.

		\subsubsection{Flow Inter-arrival Time}

		\begin{figure}
			\centering
				\includegraphics[width=0.49\textwidth]{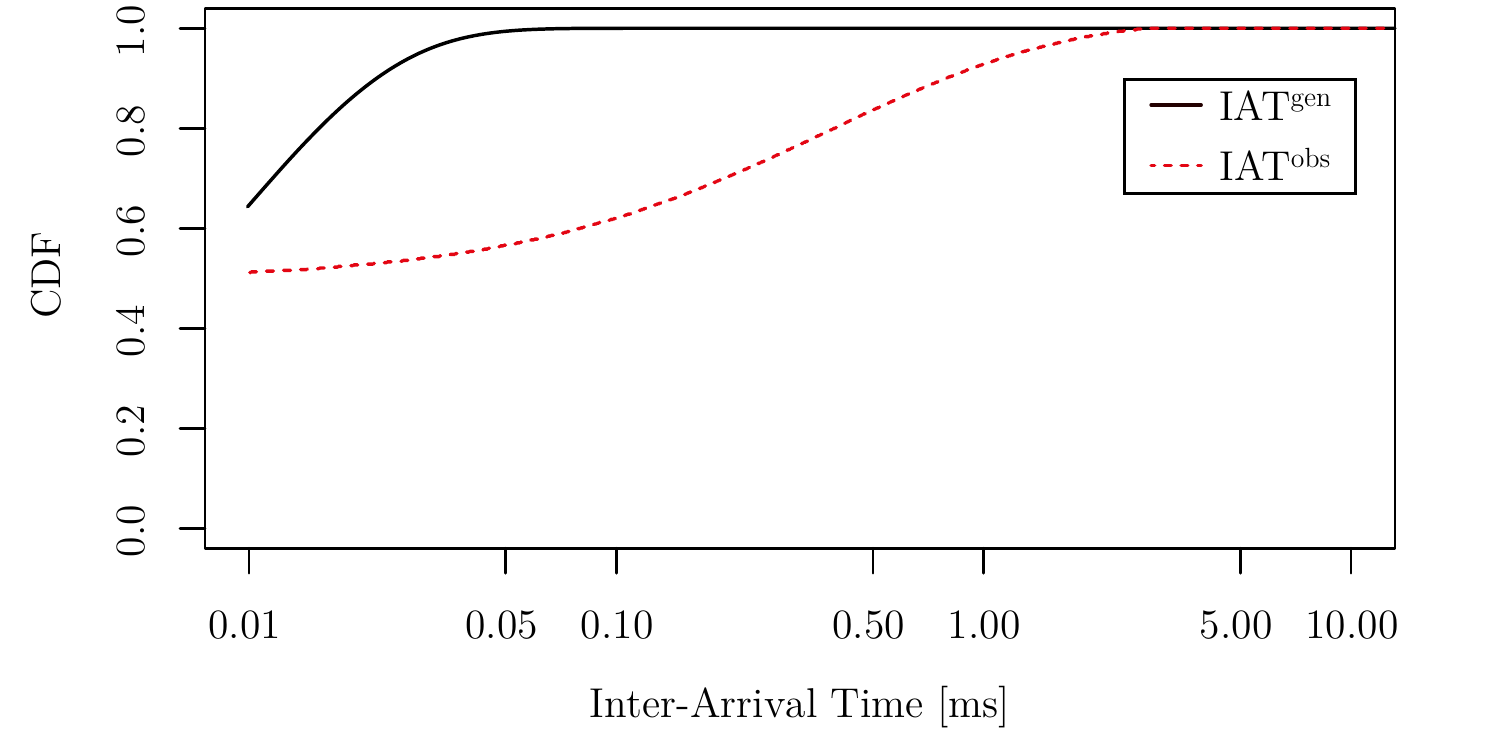}
				\caption{Comparison between \iat{obs} and \iat{gen}.}
				\label{fig:time_plot}
			%erzeugt durch die python scripte in trafficAnalyzer
		\end{figure}
		
		To compute \iat{gen}, we used the same 16 traffic schedules as before.
		In the Flowset Creator, \iat{gen} is manipulated such that the bytes contained in all generated flows
		are matching the total traffic of the traffic matrix generated in the Traffic Matrix Generator.
		Both \iat{gen} and \iat{obs} can be seen in Figure~\ref{fig:time_plot}.
		Apparently, these distributions do not match.
		The reason for this mismatch is the manipulation done in the Flowset Creator.
		With \iat{obs} extracted from \cite{MSR-datacenters} it was not possible to
		create enough flows to fill up the generated traffic matrices.
		This can have two causes: Either the
		\iat{obs} reported in \cite{MSR-datacenters} does not match the used \size{obs} or the data provided in \cite{MSR-datacenters}
		has such a low resolution that we were not able to fully recover it.
		%Another factor that we didn't take care of in the Flowset Creator are the induced ACK streams.
		%We only draw the inter-arrival time of outgoing payload flows from the given input distribution and do not include the
		%inter-arrival time of ACK flows into the flow generation process.

\section{Conclusion}
\label{sec:conclusion}	 
	The traffic generator \genname{} presented in this work creates a \lf{} traffic schedule for arbitrary sized data centers.
	When the scheduled payloads are transported using TCP, this produces Layer~2 traffic
	with properties that can be defined in advance using a set of probability distributions.
	Our evaluation showed that \genname{} reproduces these properties with high accuracy.
	Only the generated flow inter-arrival time distribution does not match our chosen target distribution.
	As \genname{} manipulates the inter-arrival time distribution to
	adjust the amount of flows to the given traffic matrices, this is not surprising.
%	Traffic is described by six probability distributions that can be chosen arbitrary. 
%	This makes our process highly versatile in a sense that traffic for various different scenarios can be generated.
	
	Given that \genname{} generates a schedule of payload transmissions between all hosts in a data center
	it is suitable for simulations, network emulations, and testbed experiments. 
	Using our generated traffic schedule combined with a large-scale network emulator such as MaxiNet,
	novel networking ideas can be evaluated under highly realistic conditions which 
	brings new ideas a step closer to deployment in production environments.

\section*{Acknowledgments}
	This work was partially supported by the German Research Foundation (DFG) within the Collaborative Research Centre ``On-The-Fly Computing'' (SFB 901).

\bibliography{trafficGenerator}{}
\bibliographystyle{unsrt}

% That's all folks!
\end{document}